\documentclass[twocolumn,a4paper,12pt]{article}
\usepackage[margin=1.3cm]{geometry}
\usepackage{comment}
\usepackage{amssymb,extarrows,graphicx,subfigure,setspace}
\usepackage{cite}
\usepackage{slashed}
\usepackage{tensor}
\usepackage[toc,page]{appendix}
\usepackage{color}
\usepackage{xcolor}
\usepackage{physics}
\usepackage{hyperref}
\usepackage{dirtytalk}
\hypersetup{colorlinks=true, linkcolor=blue, citecolor=red, linktoc=page}
\makeatother

\usepackage{float}
\usepackage{textcomp}
\usepackage{amsmath}
\newmuskip\pFqmuskip

\newcommand*\pFq[6][8]{%
  \begingroup
  \pFqmuskip=#1mu\relax
  \mathcode`\,=\string"8000
  \begingroup\lccode`\~=`\,
  \lowercase{\endgroup\let~}\pFqcomma
  {}_{#2}F_{#3}{\left[\genfrac..{0pt}{}{#4}{#5};#6\right]}%
  \endgroup
}
\newcommand{\pFqcomma}{\mskip\pFqmuskip}

\usepackage{mathrsfs}
\usepackage{enumitem}

\setlist[enumerate]{leftmargin=2.5em}
\setlist[itemize]{leftmargin=2.5em}

\newcommand{\be}{\begin{equation}}
\newcommand{\bea}{\begin{eqnarray}}
\newcommand{\eea}{\end{eqnarray}}
\newcommand{\ba}{\begin{array}}
\newcommand{\ea}{\end{array}}
\newcommand{\ee}{\end{equation}}
\newcommand{\bes}{\begin{equation*}}
\newcommand{\beas}{\begin{eqnarray*}}
\newcommand{\eeas}{\end{eqnarray*}}
\newcommand{\ees}{\end{equation*}}

\numberwithin{equation}{section}

\begin{document}
\twocolumn[
\begin{center}
\Large{\bf Can Static Black Holes in Massive Gravity Serve as Candidates for Aschenbach-Like Phenomena?}\\
\small \vspace{0.4cm}
{\bf Mohammad Ali S. Afshar $^{\star}$ $^{\dag}$ $^{\ddag}$ \footnotemark[1]}, \quad
{\bf Jafar Sadeghi $^{\star}$ $^{\dag}$ \footnotemark[2]}, \quad
{\bf Tahereh Azizi $^{\star}$\footnotemark[3]}, \quad
{\bf A. S. Sefiedgar $^{\star}$\footnotemark[4]}\\
\vspace{0.4cm}
$^{\star}$Department of Physics, Faculty of Basic Sciences,\\
University of Mazandaran, P. O. Box 47416-95447, Babolsar, Iran\\
\small \vspace{0.1cm}
$^{\dag}$School of Physics, Damghan University, P. O. Box 3671641167, Damghan, Iran\\
\small \vspace{0.1cm}
$^{\ddag}$Center for Theoretical Physics, Khazar University, 41 Mehseti Street, Baku, AZ1096, Azerbaijan\\
\small \vspace{0.1cm}
\end{center}]

\footnotetext[1]{Email:~~~m.a.s.afshar@gmail.com}
\footnotetext[2]{Email:~~~pouriya@ipm.ir}
\footnotetext[3]{Email:~~~t.azizi@umz.ac.ir}
\footnotetext[4]{Email:~~~a.sefiedgar@umz.ac.ir}
\begin{abstract}

The original Aschenbach effect refers to a non-monotonic profile of the azimuthal velocity measured by locally non-rotating observers (LNRFs) around rapidly rotating Kerr black holes; it should not be identified with non-monotonicity of the Boyer--Lindquist coordinate angular velocity. In the present work, we investigate related kinematic behavior in three static, spherically symmetric black-hole solutions in massive gravity: (i) Bardeen--AdS black holes, (ii) nonlinear charged AdS black holes, and (iii) ModMax--dRGT-like black holes. For the parameter choices considered, each spacetime admits an unstable and a stable photon sphere outside the event horizon. We examine both the coordinate angular velocity and the locally measured azimuthal speed, using the latter for a
kinematic comparison with the Kerr Aschenbach effect. For all three benchmark geometries, the locally measured speed of equatorial timelike circular geodesics has an interior minimum between the two photon-sphere radii. We also show analytically that, in these static geometries, a perfect fluid with constant specific angular momentum has a local azimuthal speed proportional to the photon-sphere potential, and we construct illustrative equilibrium-torus configurations in which the speed profile is non-monotonic within the fluid. These results establish a geometric and kinematic possibility, rather than a completed accretion-flow model: we do not calculate fluid oscillation modes, QPO frequencies, or predicted observational spectra. Accordingly, any connection with observed high-frequency quasi-periodic oscillations (QPOs) remains a question for future quantitative investigation, not a conclusion established by the present analysis.

\vspace{0.05em}
\end{abstract}
\tableofcontents
\section{Introduction}
In the strong-field regime of general relativity, orbital motion can display behavior that has no direct Newtonian counterpart \cite{1,2,3,4,13,chandrasekhar1985,frolovNovikov1998}. The Aschenbach effect, identified in rapidly rotating Kerr spacetime, is specifically the reversal of the radial gradient of the azimuthal orbital velocity measured by locally non-rotating observers (LNRFs) \cite{7,41,42,vahedi2021,bardeenPressTeukolsky1972}. This local, physically measured velocity must be distinguished from the Boyer--Lindquist coordinate angular velocity $\Omega=d\phi/dt$. For co-rotating equatorial circular geodesics, the latter is
\begin{equation}
\Omega_{\rm Kerr}(r)=\frac{\sqrt{M}}{r^{3/2}+a\sqrt{M}},
\end{equation}
whose derivative is
\begin{equation}
\frac{d\Omega_{\rm Kerr}}{dr}=-\frac{3\sqrt{Mr}}{2\left(r^{3/2}+a\sqrt{M}\right)^2}<0 \qquad (r>0).
\end{equation}
Thus, this coordinate angular velocity is monotonic and does not itself display the Aschenbach reversal. The original effect concerns the LNRF-measured azimuthal speed, whose radial behavior is affected by frame dragging and the local geometry; these quantities should not be conflated.

Accretion-disk dynamics and X-ray timing provide important context for interpreting variability from compact objects \cite{5,6,8,vanderKlis1989,balbusHawley1991,ingramDone2011}. Spectral analyses provide complementary probes of accreting black-hole systems \cite{13.1,reynoldsNowak2003}. Possible links between strong-field orbital dynamics and QPOs have also been explored in distinct, model-dependent settings. Aschenbach et al. analyzed characteristic timescales in X-ray and infrared flares from Sgr A* in terms of dynamical frequencies associated with accretion flows \cite{40}. Subsequent studies proposed that the LNRF-velocity hump in nearly extreme Kerr geometries could be related to high-frequency QPOs, including those reported for GRS 1915+105 \cite{41,42}. These proposals rely on the rotating Kerr geometry and on additional assumptions about disk oscillations; they do not imply that any non-monotonic velocity profile in an arbitrary spacetime automatically explains observed QPOs. The present static-spacetime analysis therefore treats observational relevance as a question requiring separate quantitative testing, rather than as a consequence of non-monotonicity alone.

The dependence of the original Kerr effect on frame dragging motivates a related but distinct question: can a static spacetime support a non-monotonic profile of locally measured orbital speed? In our previous work \cite{14}, we investigated analogous kinematic behavior in non-rotating black-hole geometries. In the present paper, the coordinate angular velocity $\Omega(r)$ is also examined, but it is not, by itself, the defining quantity of the original Aschenbach effect. To make the comparison physically explicit, we additionally evaluate the azimuthal speed measured by static observers and, below, the speed profile of constant-specific-angular-momentum fluid configurations. In static spherical geometries, the existence of both an unstable and a stable photon sphere provides a natural interval in which the locally measured speed of equatorial circular geodesics can be non-monotonic.

Stable photon spheres outside an event horizon are absent in many familiar static black-hole solutions, but they can occur in particular nonstandard geometries, including some models discussed in Refs. \cite{15,15.1,16,17,18,19}. Closely related topological analyses have been applied to static black holes in effective quantum gravity: Liu, Wu, and Wang found unstable light rings in two classes of solutions \cite{eqg2024}, while their subsequent study of a third solution identified stable and unstable light-ring branches \cite{eqg2025}. These studies provide relevant methodological context, although they concern effective-quantum-gravity geometries rather than the massive-gravity solutions considered here. The present work examines three specific massive-gravity metrics and the relation between their photon-sphere structure and timelike circular-orbit kinematics. The existence of a stable photon sphere is a geometric property; by itself, it does not establish an observational signature. Any astrophysical interpretation requires additional modeling.

Motivated by previous studies of static black-hole geometries in modified-gravity settings \cite{14,19}, we examine whether three specific spherically symmetric massive-gravity solutions admit the geometric and kinematic conditions for Aschenbach-like behavior: (i) Bardeen--AdS black holes, (ii) nonlinear charged AdS black holes, and (iii) ModMax--dRGT-like black holes. The analysis concerns these selected solutions and parameter choices; it does not establish a generic property of massive-gravity black holes.

For each spacetime, we determine parameter regimes in which a stable photon sphere coexists with an event horizon, examine the effective-potential structure, and analyze equatorial timelike circular orbits. We distinguish the original Kerr effect, defined through a locally measured azimuthal velocity and associated in the Kerr case with frame dragging, from related non-monotonic behavior in static geometries. We also test whether constant-specific-angular-momentum fluid tori can contain the relevant velocity minimum. The results establish geometric and kinematic conditions that may motivate future strong-field studies of massive-gravity models; they do not, without a fluid-oscillation model and frequency calculation, establish an observational explanation for any particular QPO signal.

\section{Methodology}

To analyze the photon sphere structure, we adopt the topological charge-based approach developed in \cite{20,21} and further applied in our previous studies \cite{15,16}. The stability criteria for timelike circular orbits (TCOs) are evaluated using the standard orbital relations established in \cite{22,23,24}.

\subsection{Topological Photon Sphere} 

We consider a spherically symmetric four-dimensional spacetime with the metric:
\begin{equation}\label{(1)}
\begin{split}
&ds^{2}=-f(r)dt^{2} + \frac{dr^{2}}{g(r)} + h(r)(d\theta^{2}+\sin^{2}\theta d\varphi^{2}) = \\ 
& g_{tt}dt^{2} + g_{rr}dr^{2} + g_{\theta\theta}d\theta^{2} + g_{\varphi\varphi}d\varphi^{2}.
\end{split}
\end{equation}
The effective potential for the photon sphere, denoted by $H(r, \theta)$, is defined as:
\begin{equation}\label{(2)}
\begin{split}
H(r,\theta)=\sqrt{\frac{-g_{tt}}{g_{\varphi\varphi}}}=\frac{1}{\sin\theta}\bigg(\frac{f(r)}{h(r)}\bigg)^{1/2}.
\end{split}
\end{equation}
To determine the topological charges, we construct a two-dimensional vector field $\phi=(\phi^{r}, \phi^{\Theta})$ \cite{20,21}:
\begin{equation}\label{eq:topological-vector-field}
\phi=(\phi^{r}, \phi^{\Theta}),
\end{equation}
which is normalized as $n^\iota = \phi^\iota / \|\phi\|$ where $\iota=1,2$ and $\|\phi\|=\sqrt{\phi\cdot\phi}$. The components of this vector field in terms of the potential $H$ are given by:
\begin{equation}\label{(5)}
\begin{split}
&\phi^r = \frac{\partial_r H}{\sqrt{g_{rr}}} = \sqrt{g(r)}\partial_{r}H,\\
&\phi^\Theta = \frac{\partial_\theta H}{\sqrt{g_{\theta\theta}}} = \frac{\partial_\theta H}{\sqrt{h(r)}}.
\end{split}
\end{equation}
The topological charge associated with each photon sphere is assigned via the winding number of the vector field $\phi$ \cite{15,16,21}. Topological methods have also been used to classify light rings in other static modified-gravity geometries. In particular, Ref. \cite{eqg2024} found unstable light rings in two effective-quantum-gravity solutions, whereas Ref. \cite{eqg2025} reported both stable and unstable branches for another static solution. Here we apply the corresponding static, spherically symmetric construction to three massive-gravity metrics and combine it with an analysis of timelike circular orbits. For the benchmark black-hole configurations studied here, the effective potential has an unstable photon sphere at a local maximum and a stable photon sphere at a local minimum. Their existence and stability are determined for each solution and parameter set rather than inferred from a general distinction between horizon-enclosed and horizonless geometries.
\subsection{Time-like Circular Orbits (TCOs)}

The geodesic structure in any gravitational background is governed by the effective potential derived from the underlying action. The radial motion of null and timelike geodesics can be
described by an effective potential. Circular orbits satisfy the conditions of vanishing radial velocity and radial acceleration. Their radial stability is determined by perturbing the orbit while holding the conserved quantities fixed.\\ 
By spherical symmetry, the orbital plane can be chosen as the equatorial plane without loss of generality. Using the metric in Eq.~(\ref{(1)}), we define the following quantities
\cite{22}:
\begin{equation}\label{(6)}
A = g_{\varphi \varphi} E^{2} + g_{tt} L^{2},
\end{equation}
\begin{equation}\label{(7)}
B = -g_{\varphi \varphi} g_{tt},
\end{equation}
where $E$ and $L$ represent the particle's energy and angular momentum, respectively. For the metric in Eq. (\ref{(1)}), the normalization of the four-velocity (or null tangent) gives \cite{22}:

\begin{equation}\label{(8)}
2 \mathfrak{L} = -\frac{A}{B}+\frac{\dot r^{\,2}}{g(r)} = \varrho,
\end{equation}
where $\varrho=-1$ and $0$ correspond to timelike and null geodesics, respectively. For a circular orbit, $\dot r=0$, so Eq. (\ref{(8)}) reduces to $-A/B=\varrho$. The effective potential for radial motion is expressed as:
\begin{equation}\label{(9)}
V_{\text{eff}}(\varrho, r) = \varrho + \frac{A}{B}.
\end{equation}
The angular velocity $\Omega$ and the parameter $\beta$ are given in terms of the metric components by \cite{22}:
\begin{equation}\label{(10)}
\Omega = -\frac{g_{tt} L}{g_{\varphi \varphi} E},
\end{equation}
\begin{equation}\label{(11)}
\beta = -g_{tt} - 2 g_{t\varphi} \Omega - g_{\varphi\varphi} \Omega^2 = -r^{2} \Omega^{2} + f(r).
\end{equation}
Physical realizability of TCOs requires $\beta>0$ and $\Omega^2>0$. For an equatorial circular geodesic in the static, spherically symmetric metrics considered here, the radial geodesic equation gives the explicit relations

\begin{equation}\label{eq:circular-geodesic-omega}
\Omega^2=\frac{f'(r)}{2r},
\qquad
\beta=f(r)-\frac{r f'(r)}{2}.
\end{equation}
Differentiating these expressions yields
\begin{equation}\label{eq:beta-omega-derivative}
\frac{d\beta}{dr}=-r^2\frac{d(\Omega^2)}{dr}.
\end{equation}
At an extremum of a positive circular-orbit angular velocity, $d\Omega/dr=0$ and $d\beta/dr=0$. At such a point, $d^2\beta/dr^2=-r^2d^2(\Omega^2)/dr^2$, so a local minimum of $\Omega$ corresponds to a local maximum of $\beta$, and vice versa.

In regions where $\beta<0$, no timelike circular orbit exists; such branches, even if $\Omega^2$ remains positive as a formal expression, are not physically admissible TCOs. For the three benchmark parameter sets analyzed below, the physical circular-orbit domain outside the event horizon is the open interval $r_{\rm U}<r<r_{\rm S}$ between the two photon-sphere radii. At these endpoints $\beta=0$ and the circular orbits are null, so the endpoints themselves are excluded from the timelike domain. The angular-velocity plots are therefore restricted to this physical interval. Further details on orbit stability can be found in Refs. \cite{22,23,24}.

\section{Black Holes in Massive Gravity: Theoretical Foundations}

The historical development of massive gravity traces back to the linear formulation by Fierz and Pauli for a massive spin-2 field in flat spacetime \cite{26}. However, this linearized theory suffers from the van Dam--Veltman--Zakharov (vDVZ) discontinuity, failing to reproduce General Relativity (GR) in the massless limit \cite{27}. This issue was partially addressed by Vainshtein, who showed that nonlinear interactions can generate strong-field screening effects ("Vainshtein screening") that suppress the helicity-0 mode near massive sources \cite{28}. Early nonlinear formulations also faced the Boulware--Deser ghost problem \cite{29}. 

A major development was the de Rham--Gabadadze--Tolley (dRGT) construction of a nonlinear massive-gravity theory with a specially tuned potential \cite{derhamGabadadze2010,dRGT2011,25}. The absence of the Boulware--Deser degree of freedom in the relevant formulation was established in the Hamiltonian analysis by Hassan and Rosen \cite{30}.

The formulation of dRGT massive gravity opened new avenues for exact black-hole solutions and related applications, including holographic and cosmological settings \cite{31,volkov2012,chamseddineVolkov2011}. The solutions relevant to the present discussion broadly include:
\begin{itemize}
    \item \textbf{Schwarzschild-like Solutions:} Configurations that preserve spherical symmetry while embedding the graviton mass parameters, which act as effective modifications to the asymptotic geometry or mimic cosmological terms.
    \item \textbf{Hairy Black Holes:} Solutions endowed with primary "hair" arising from the non-trivial profile of Stückelberg fields or reference metric interactions, evading standard no-hair theorems of General Relativity.
    \item \textbf{Bimetric Extensions:} Frameworks involving two interacting metrics (one dynamical and one reference, which can also be promoted to dynamical), with rotating black-hole solutions constructed in particular formulations \cite{ayonbeatoBigravity2016,babichevFabbri2014}.
\end{itemize}

The presence of a graviton mass fundamentally alters gravitational dynamics and collapse processes. The modified field equations are symbolically expressed as:
\begin{equation}\label{eq:massive-gravity-field}
G_{\mu\nu} = 8\pi G T_{\mu\nu}^{\text{(matter)}} + T_{\mu\nu}^{\text{(mass)}},
\end{equation}
where the effective mass tensor $T_{\mu\nu}^{\text{(mass)}}$ depends non-trivially on the metric and reference structures. This modification introduces several distinct physical characteristics:
\begin{enumerate}
    \item \textbf{Altered Propagation and Asymptotics:} Gravitational fluctuations experience Yukawa-like suppression at large distances, modifying the large-scale structure of spacetime.
    \item \textbf{The Vainshtein Mechanism:} Active in high-density or strong-field regimes near the horizon, this mechanism dynamically suppresses non-GR polarizations, ensuring local consistency with standard gravity.
    \item \textbf{Hair Formation:} Boundary conditions enforced by the potential terms and Stückelberg fields require additional parameters beyond mass, charge, and spin to characterize the horizon geometry.
\end{enumerate}

While massive gravity provides compelling theoretical frameworks, offering potential insights into dark energy, singularity resolution via energy-condition violations, and novel strong-field signatures, it also introduces notable challenges. These include analytical complexity, stability and causality concerns (such as potential superluminal propagation of fluctuations), fine-tuning requirements for the small graviton mass, and stringent constraints from gravitational wave observations.

\subsection{Bardeen-AdS Black Holes in Massive Gravity}

The action for a Bardeen black hole coupled to nonlinear electrodynamics in massive gravity with a negative cosmological constant $\Lambda=-3/l^2$ is considered in Refs. \cite{33,34}:

\begin{equation}\label{(12)}
\begin{split}
S = \frac{1}{2}\int d^{4}x\sqrt{-g}\Bigg[&R-2\Lambda\\
&+m_{g}^{2}\sum_{i=1}^{4}c_{i}\mathcal{U}_{i}(g,f)\\
&-\frac{1}{4\pi}\mathcal{L}(F)\Bigg].
\end{split}
\end{equation}

where $m_g$ denotes the graviton mass, $f$ is the reference metric, and $\mathcal{U}_i$ are the symmetric polynomials of the eigenvalues of the matrix $\mathcal{K}^\mu_\nu = \sqrt{g^{\mu\alpha}f_{\alpha\nu}}$ \cite{32}. The Bardeen Lagrangian density is given by $\mathcal{L}(F) = \frac{3}{2sq^2} \left( \frac{\sqrt{2q^2F}}{1 + \sqrt{2q^2F}} \right)^{5/2}$, with $s = q/2M$ \cite{33,34}. The metric function $f(r)$ for this static, spherically symmetric spacetime is \cite{34}:
\begin{equation}
f(r) = 1 - \frac{2M r^2}{(r^{2}+q^{2})^{3/2}} + \frac{r^2}{l^2} + m_{g}^2 (C^2c_2 + \frac{Cc_1 r}{2}).
\label{eq:bardeen-metric}
\end{equation}

\begin{figure}
\begin{center}
\subfigure[]{\includegraphics[width=0.8\linewidth,height=3.3cm]{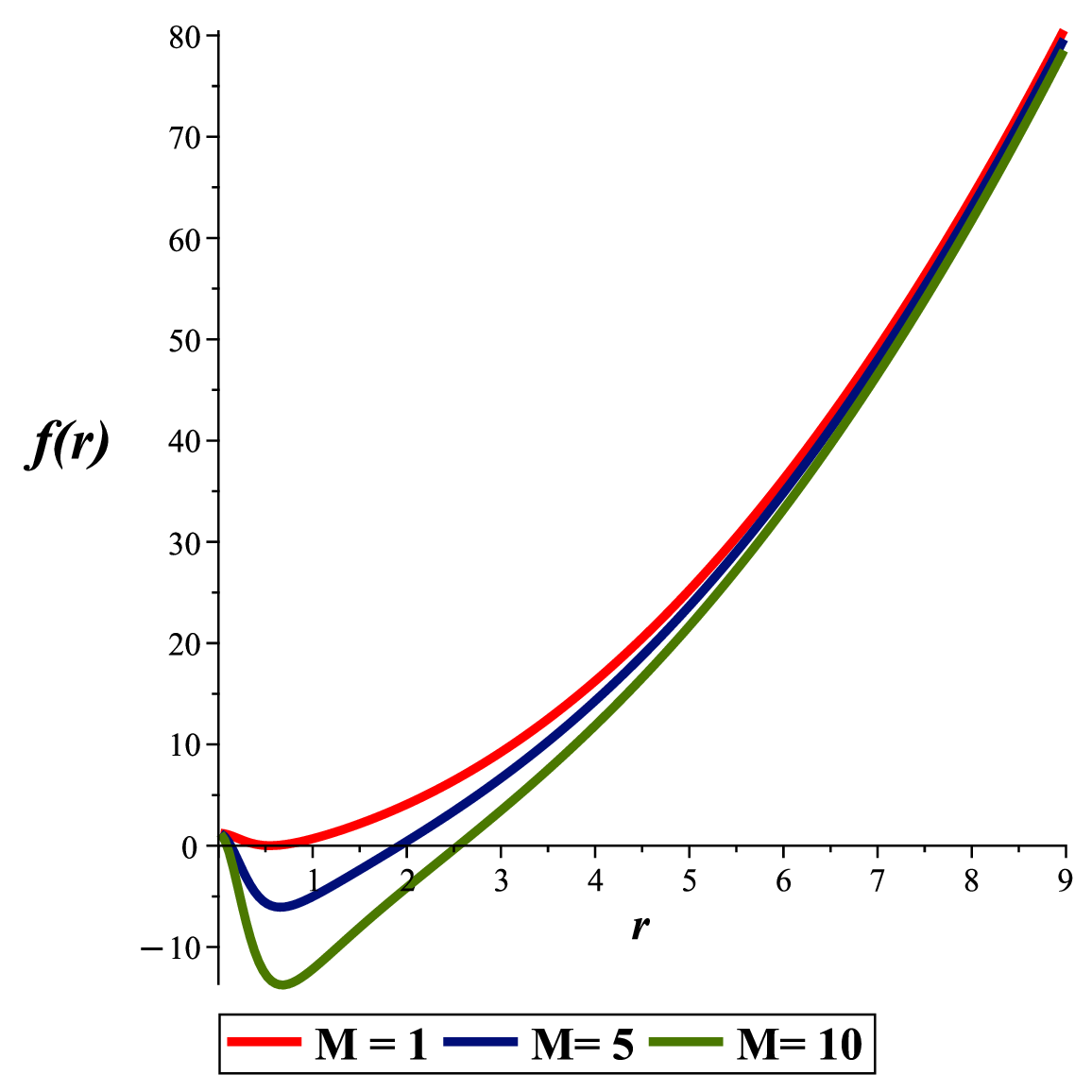}\label{1a}}\\
\subfigure[]{\includegraphics[width=0.8\linewidth,height=3.3cm]{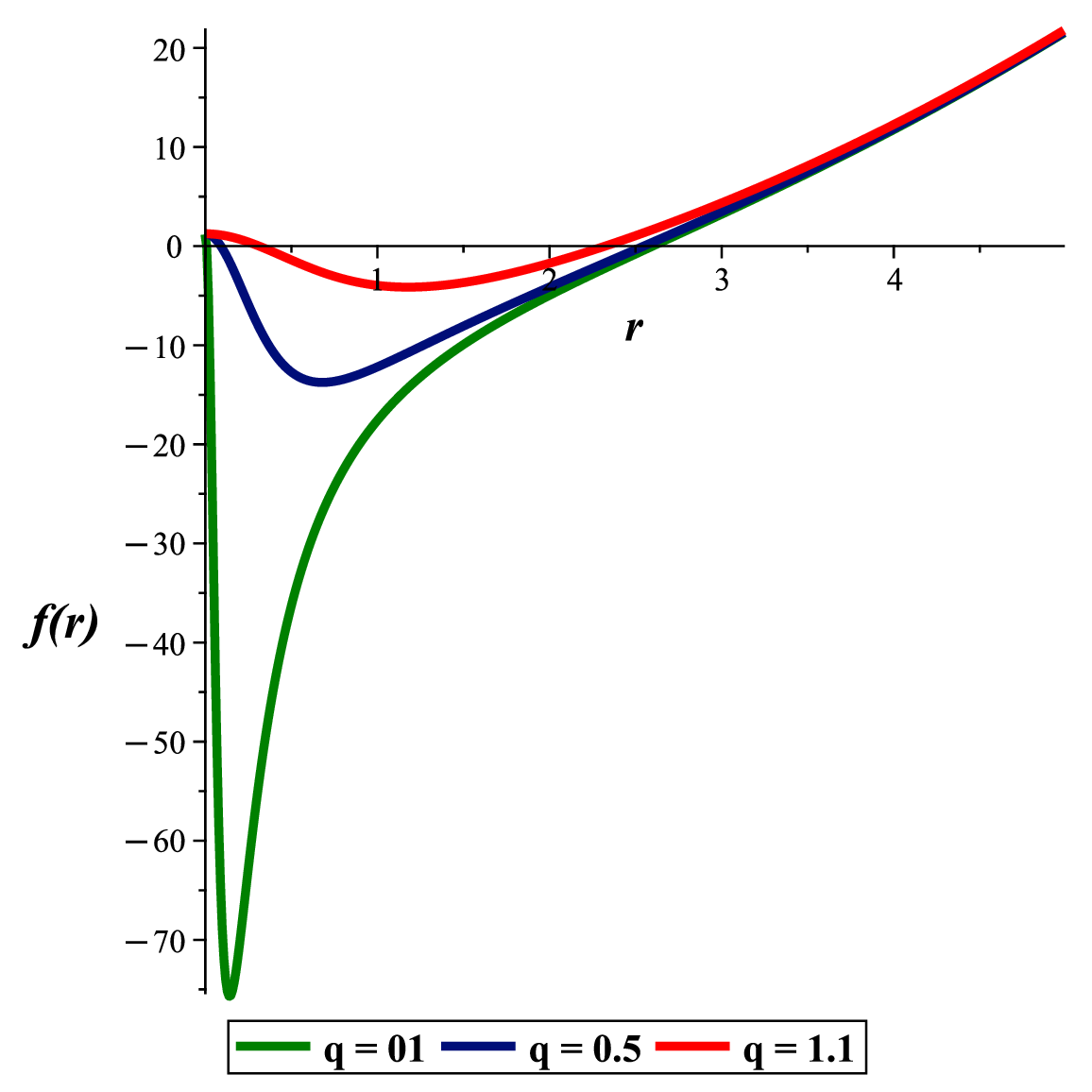}\label{1b}}\\
\subfigure[]{\includegraphics[width=0.8\linewidth,height=3.3cm]{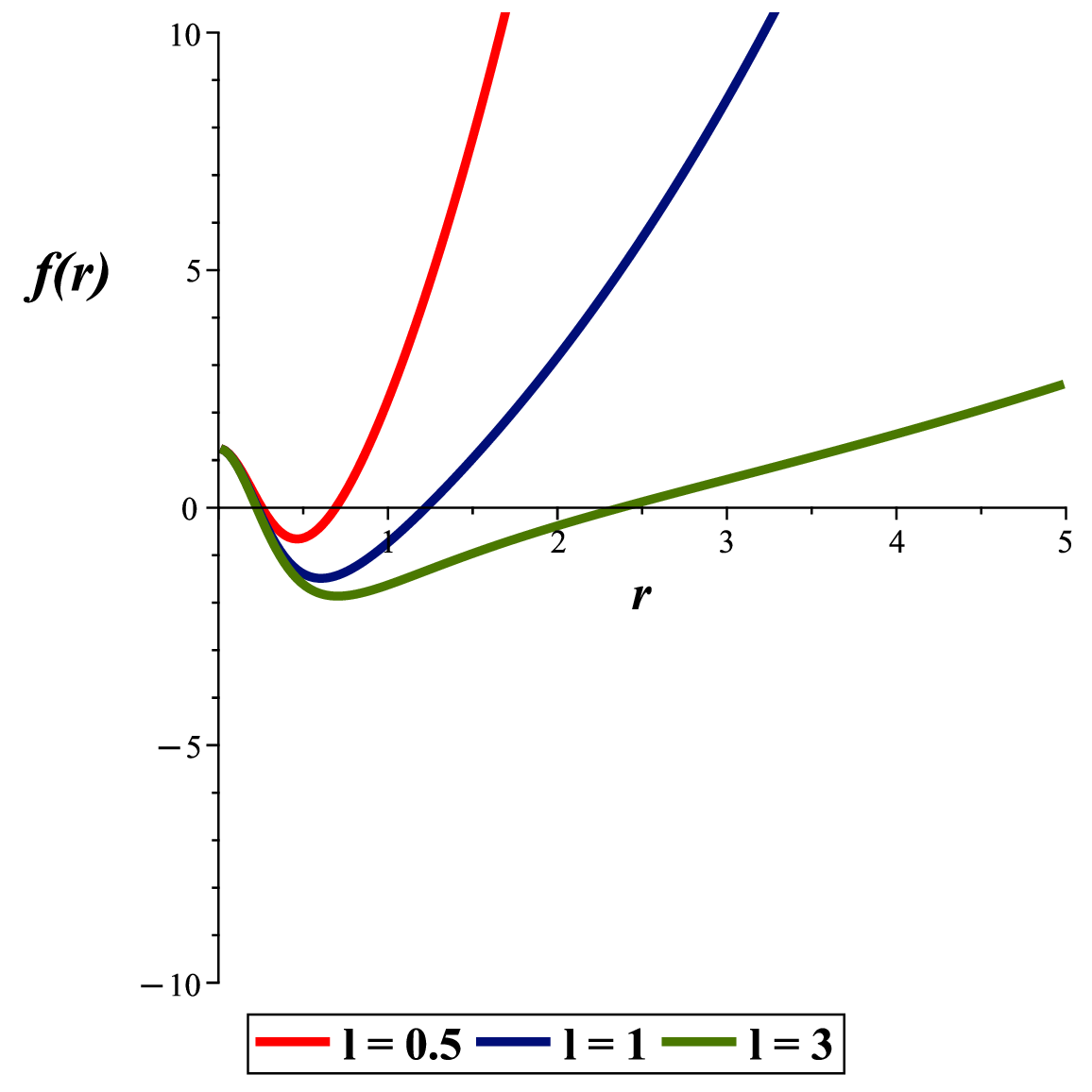}\label{1c}}\\
\subfigure[]{\includegraphics[width=0.8\linewidth,height=3.3cm]{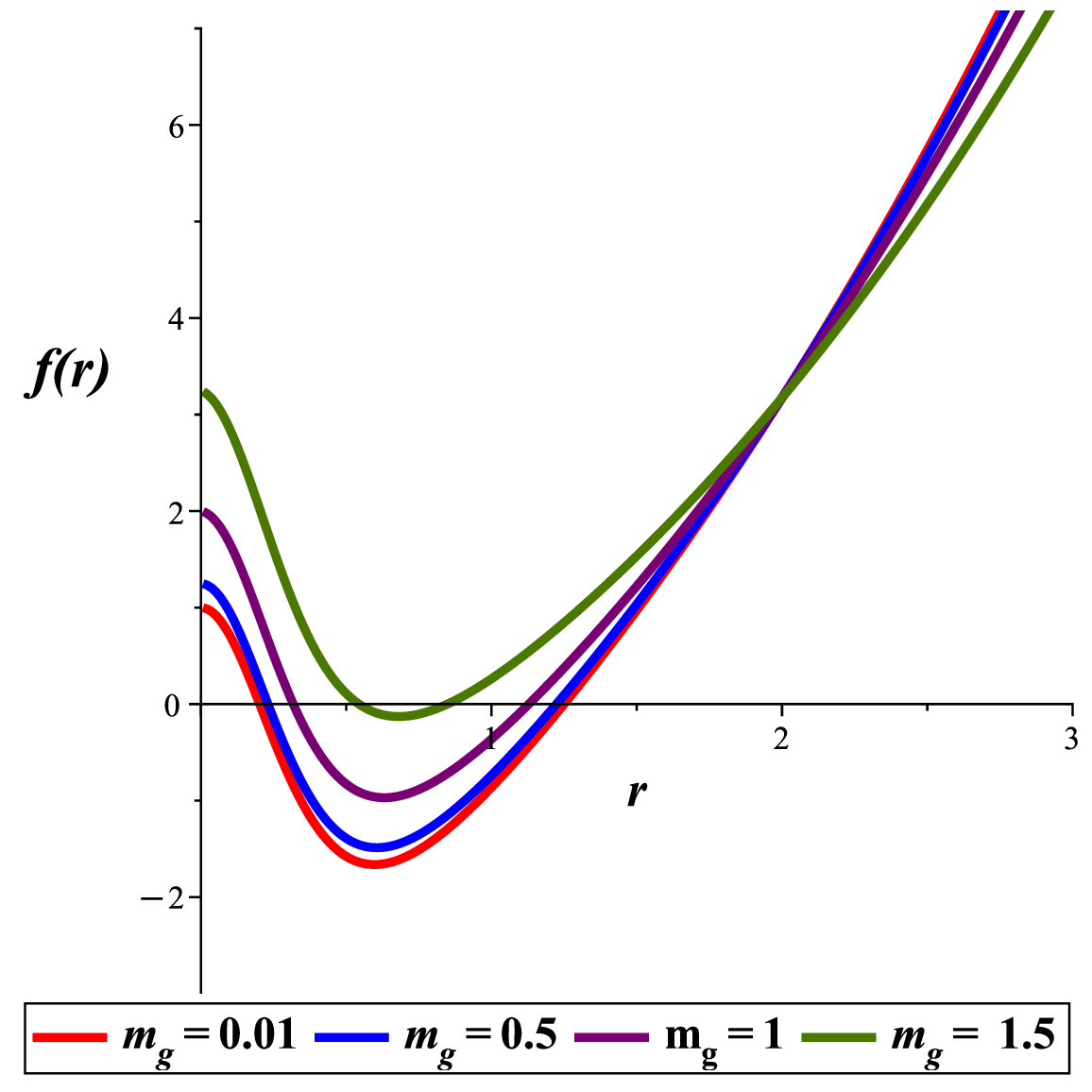}\label{1d}}\\
\subfigure[]{\includegraphics[width=0.8\linewidth,height=3.3cm]{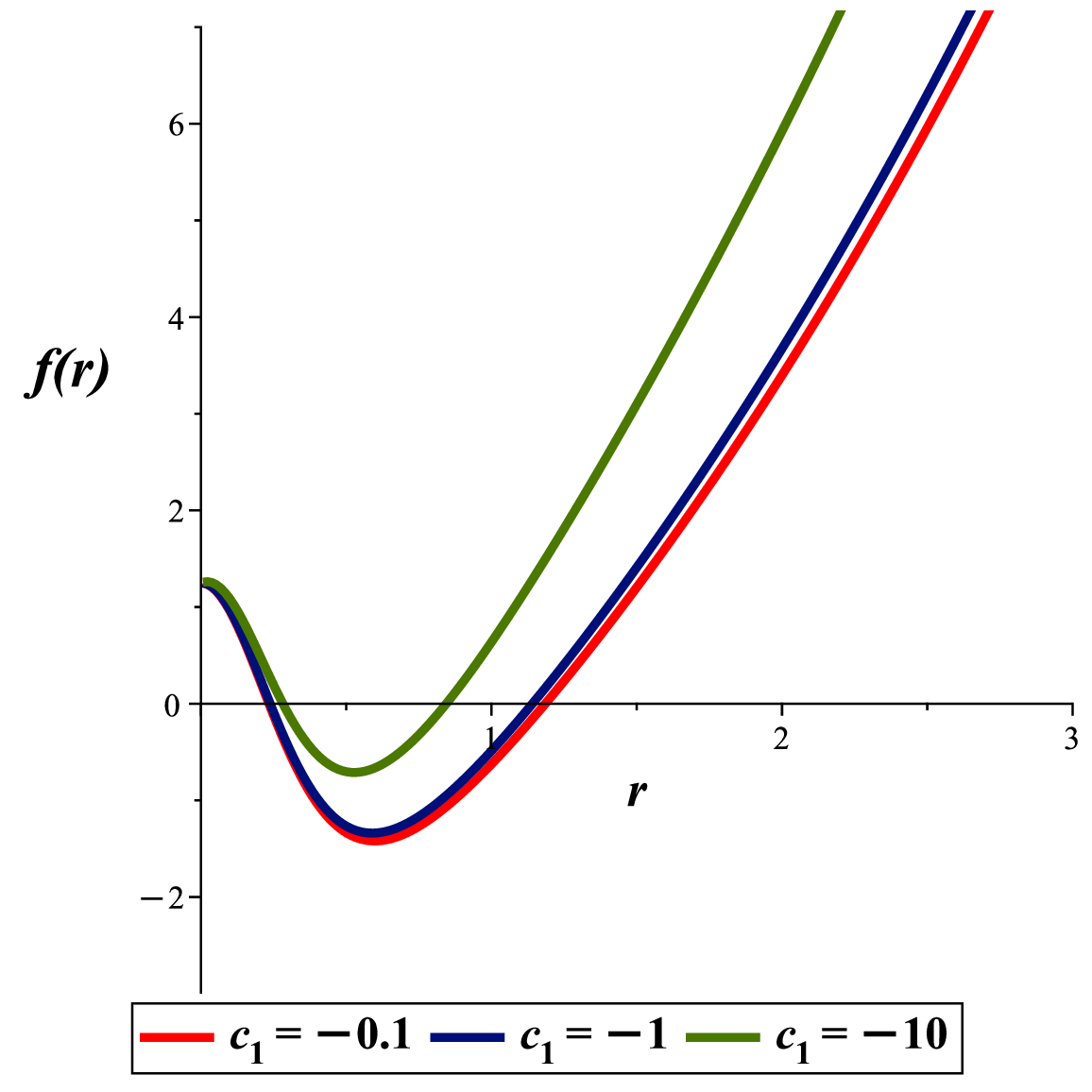}\label{1e}}
\caption{Metric function $f(r)$ for Bardeen-AdS black holes in massive gravity. Parameters are set to $l=1, C=1, c_1=-1, c_2=1, m_g=0.5$ unless otherwise specified: (a) $q=0.5$ with varying $M$; (b) $M=10$ with varying $q$; (c) $M=2, q=0.5, m_g=0.5$ with varying $l$; (d) $l=1$ with varying $m_g$; and (e) $l=1, m_g=0.5$ with varying $c_1$.}
\label{1}
\end{center}
\end{figure}

Related Bardeen-type solutions have also been studied in de Sitter and higher-curvature settings \cite{35,36,37}. The structural properties of the horizon in the present model depend sensitively on the interplay between the gravitational, electromagnetic, and massive-gravity parameters. Fig. (\ref{1}) illustrates the variation of the metric function $f(r)$ under different parametric regimes. Panels (\ref{1a}) and (\ref{1d}) demonstrate that when the black hole mass $M$ and the graviton mass $m_g$ are comparable, the system approaches a naked singularity, resulting in the disappearance of the event horizon. A similar transition is observed in Fig. (\ref{1b}), where increasing the magnetic monopole charge $q$ forces the system toward a singular state. The influence of the AdS radius $l$ is highlighted in Fig. (\ref{1c}); while larger $l$ values generally shift the horizon radius, the changes shown for larger values (e.g., $l=20$) concern the number and locations of roots of the metric function. Such a change in horizon-root structure is not, by itself, evidence of a dynamical instability of the outer horizon. Finally, the parameter $c_1$ exerts a negligible effect in small domains, but beyond a critical value, it significantly reduces the inter-horizon region by driving the Cauchy and event horizons toward each other, as shown in Fig. (\ref{1e}).

\subsubsection{In Search of the Aschenbach-like Phenomenon}

Given the complexity of the model and its parameter space, an analytical treatment is generally intractable, necessitating a numerical approach. We set a baseline configuration by assigning the parameters $M = 2$, $l = 1$, $q = 0.5$, $C = 1$, $c_{1} = -1$, $c_{2} = 1$, and $m_{g} = 0.5$. Numerical solution of the metric function reveals two distinct positive roots for this parameter set: an inner (Cauchy) horizon at $r = 0.2306$ and an event horizon at $r = 1.2232$. This identifies the horizon structure of the selected solution; the existence of these horizons alone does not establish the Weak Cosmic Censorship Conjecture (WCCC).

\begin{figure}[H]
\begin{center}
\subfigure[]{\includegraphics[width=0.8\linewidth,height=8cm]{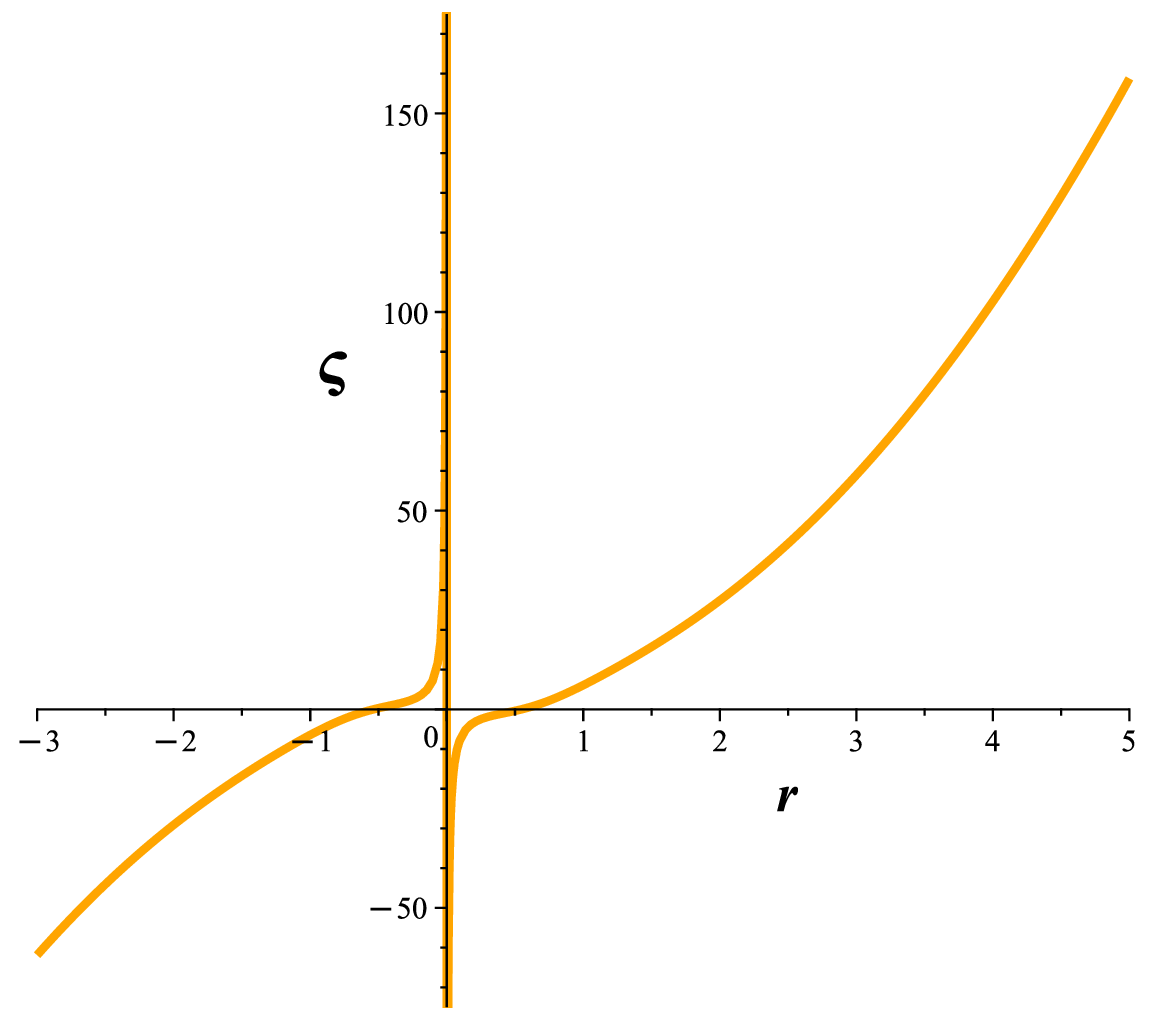}}
\caption{\small{Heat-capacity diagnostic $\varsigma(r_h)$ for
Bardeen--AdS black holes in massive gravity, evaluated along
the horizon family at fixed $q=0.5$, $l=1$, $C=1$, $c_1=-1$,
$c_2=1$, and $m_g=0.5$. The mass varies with the horizon
radius according to Eq.~(\ref{(14)}). The benchmark
configuration has $r_h\simeq1.223287$ and $M\simeq2$.}}
\label{2}
\end{center}
\end{figure}

The mass and temperature of the black hole are given by
\begin{equation}\label{(14)}
\begin{split}
&M =\\ & \frac{\left(q^{2}+r^{2}\right)^{\frac{3}{2}} \left(2 C^{2} l^{2} c_{2} m_{g}^{2}+C \,l^{2} r c_{1} m_{g}^{2}+2 l^{2}+2 r^{2}\right)}{4 r^{2} l^{2}},\\
\end{split}
\end{equation}
\begin{equation}\label{(15)}
\begin{split}
&T =\frac{1}{8 r \,l^{2} \left(q^{2}+r^{2}\right) \pi}\\ &\Big[\left(2 C \,r^{3} c_{1} m_{g}^{2}+\left(2 C^{2} c_{2} m_{g}^{2}+2\right) r^{2}\right.\\
&\left.-C \,q^{2} r c_{1} m_{g}^{2}-4 q^{2} \left(C^{2} c_{2} m_{g}^{2}+1\right)\right) l^{2}+6 r^{4}\Big]\\
\end{split}
\end{equation}
At fixed $q$, $l$, $C$, $c_1$, $c_2$, and $m_g$, we define the heat-capacity diagnostic along the horizon family by:
\begin{equation}
\varsigma(r_h)
=
\left(\frac{\partial M}{\partial T}\right)_
{q,l,C,c_1,c_2,m_g}
=
\frac{dM(r_h)/dr_h}{dT(r_h)/dr_h}.
\label{eq:heat-capacity}
\end{equation}
Figure~(\ref{2}) displays $\varsigma(r_h)$ for the parameter set $q=0.5$, $l=1$, $C=1$, $c_1=-1$, $c_2=1$, and $m_g=0.5$. The benchmark configuration has an event horizon at $r_h\simeq1.223287$, corresponding to $M\simeq2$. Within the radial interval displayed in the figure, $\varsigma(r_h)$ is positive; therefore, the corresponding horizon solutions are locally thermodynamically stable. 

To analyze the geodesic structure, we utilize the metric coefficients in Eq. (\ref{(1)}) to derive the effective potential $H(r, \theta)$ and the components of the vector field $\phi(r, \theta)$ (Eq. (\ref{(2)}) and Eq. (\ref{(5)})):
\begin{multline}\label{(17a)}
H =\frac{1}{\sin \! \left(\theta \right) r}\Bigg[1-\frac{2 r^{2} M}{\left(q^{2}+r^{2}\right)^{\frac{3}{2}}}+\frac{r^{2}}{l^{2}}\\
+\left(\frac{1}{2} r c_{1} C +C^{2} c_{2}\right) m_{g}^{2}\Bigg]^{1/2},
\end{multline}
\begin{multline}\label{(17b)}
\phi^{r}= -\frac{\csc \! \left(\theta \right)}{\left(q^{2}+r^{2}\right)^{\frac{5}{2}} r^{2}}\\ \Bigg[\left(c_{2} m_{g}^{2} C^{2}+\frac{1}{4} r c_{1} m_{g}^{2} C +1\right)\\
\times\left(q^{2}+r^{2}\right)^{\frac{5}{2}}-3 M \,r^{4}\Bigg],
\end{multline}
\begin{multline}\label{(17c)}
\phi^{\Theta}=-\frac{\cot \! \left(\theta \right) \csc \! \left(\theta \right)}{r^{2}}\\ \Bigg[\frac{\left(\left(c_{2} m_{g}^{2} C^{2}+\frac{1}{2} r c_{1} m_{g}^{2} C +1\right) l^{2}+r^{2}\right)}{\left(q^{2}+r^{2}\right)^{\frac{3}{2}} l^{2}}\\
\times\left(q^{2}+r^{2}\right)^{\frac{3}{2}}-2 r^{2} M \,l^{2}\Bigg]^{1/2}.
\end{multline}

\begin{figure}[H]
\begin{center}
\subfigure[]{\includegraphics[width=1\linewidth,height=5.7cm]{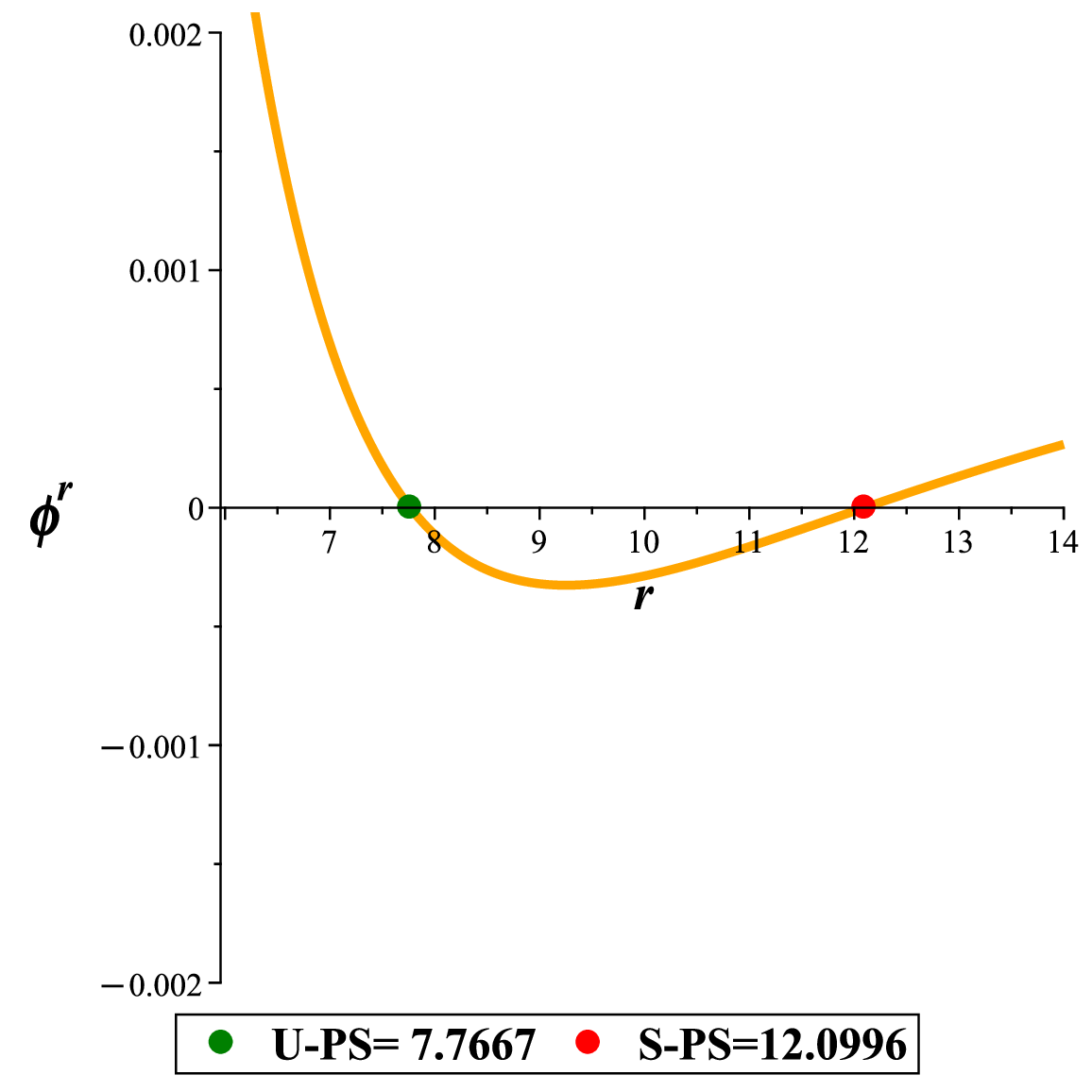}\label{3a}}\\
\subfigure[]{\includegraphics[width=1\linewidth,height=5.7cm]{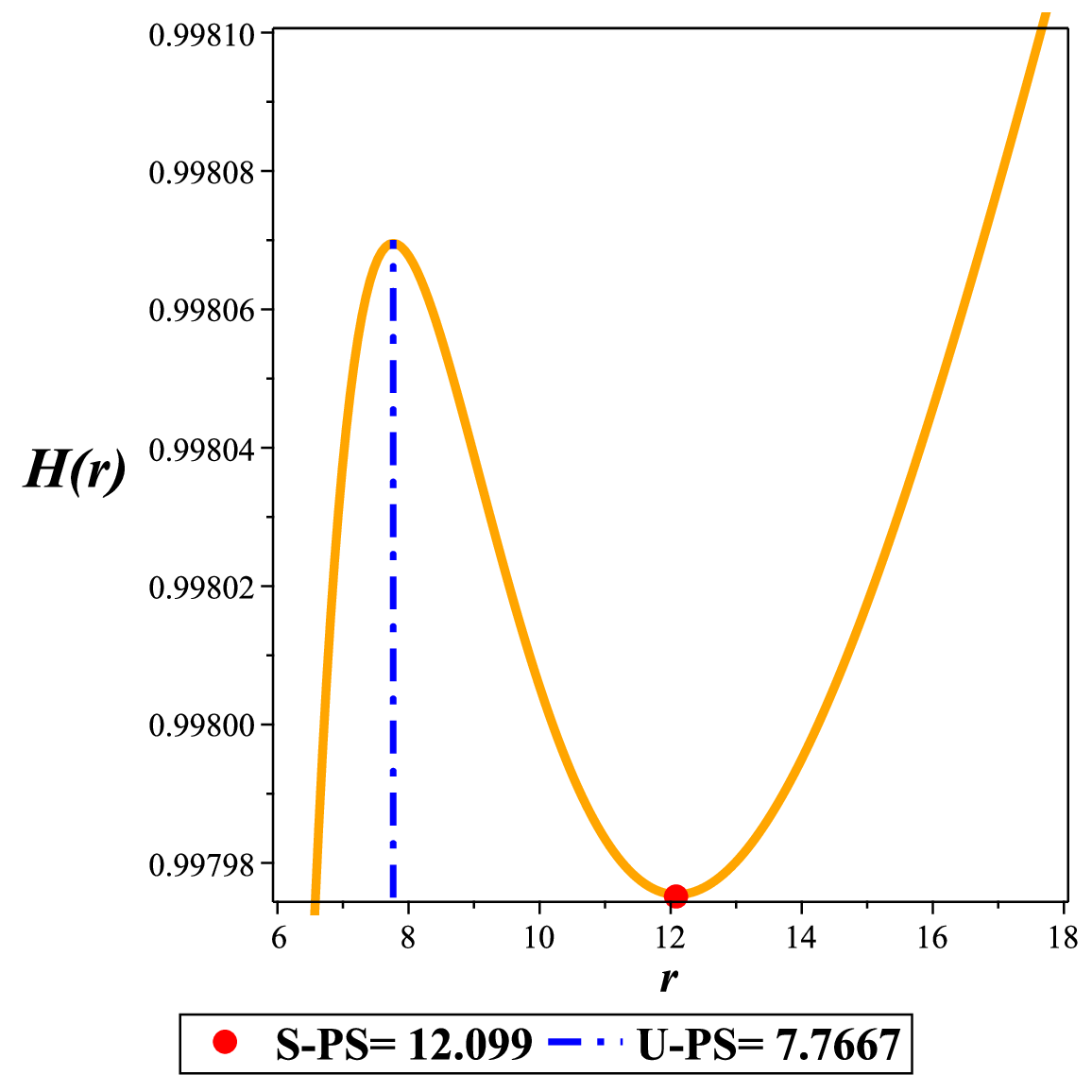}\label{3b}}
\caption{\small{(3a): The photon sphere locations at $(r, \theta) = (7.7667, 1.57)$ and $(r, \theta) = (12.0996, 1.57)$ for $M = 2, l = 1, q = 0.5, C = 1, c_{1} = -1, c_{2} = 1, m_{g} = 0.5$ in the $(r-\theta)$ plane of the normal vector field $n$; (3b): the topological potential $H(r)$ for Bardeen black holes in massive gravity.}}
\label{3}
\end{center}
\end{figure}

As illustrated in Fig. (\ref{3}), the topology of the vector field reveals the simultaneous presence of two distinct photon spheres outside the event horizon: an unstable photon sphere (U-PS, corresponding to a maximum of $H$) and a stable photon sphere (S-PS, corresponding to a local minimum of $H$).

Following the framework established by Wei and our previous work, the angular velocity $\Omega$, energy $E$, and angular momentum $L$ of TCOs are governed by \cite{14,18}:
\begin{equation}
\begin{split}
&\beta = -g_{tt} - 2 g_{t\varphi} \Omega - g_{\varphi\varphi} \Omega^2 = -r^{2} \Omega^{2}+f(r), \\
&E = \frac{f}{\sqrt{\beta}}, \quad L = \frac{r^{2} \Omega}{\sqrt{\beta}}, \\
&\Omega = \sqrt{\frac{f(r)-\beta}{r^{2}}} = \frac{L f}{E r^{2}}.
\label{(18)}
\end{split}
\end{equation}
The conditions $\beta>0$ and $\Omega^2>0$ identify the physically admissible domain of timelike circular orbits. Radial stability is assessed separately using the second derivative of the effective potential. Only roots of the marginal-stability condition that lie within the admissible domain represent physical marginally stable circular orbits(MSCOs).

\begin{figure}[H]
\begin{center}
\subfigure[]{\includegraphics[width=1\linewidth,height=6.7cm]{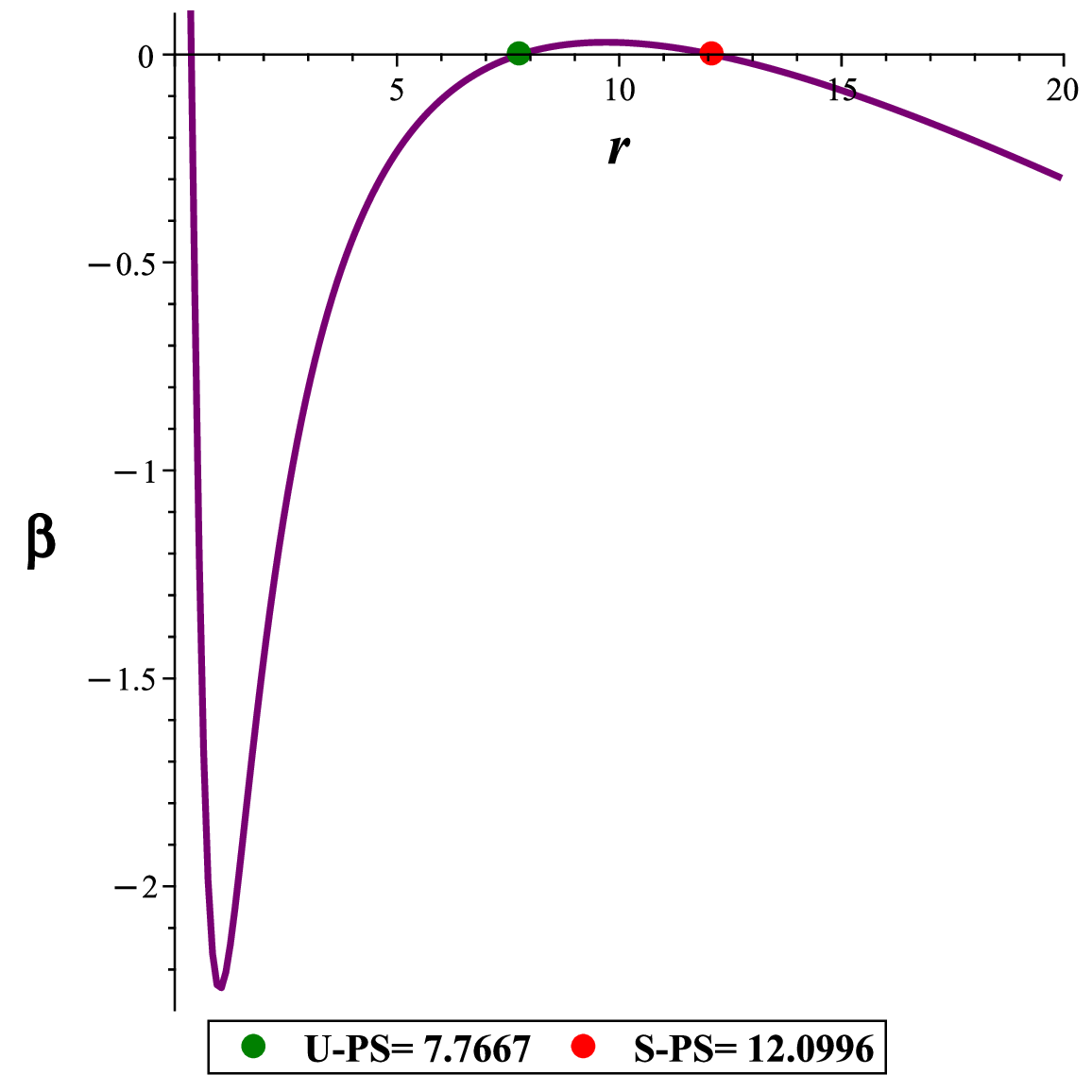}\label{4a}}\\
\subfigure[]{\includegraphics[width=1\linewidth,height=6.7cm]{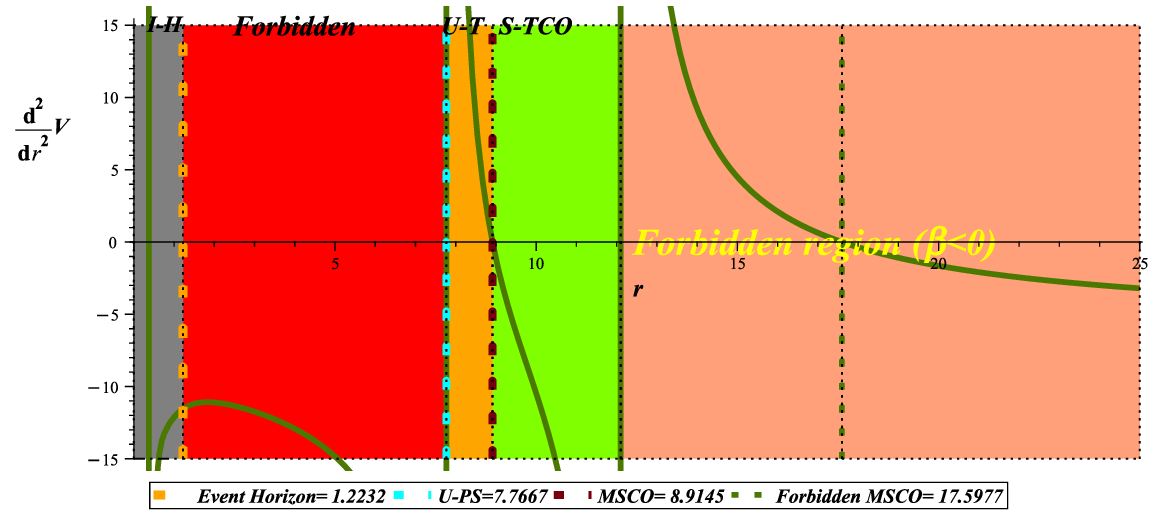}\label{4b}}
\caption{\small{(4a): $\beta$ diagram; (4b): Roots of the marginal-stability condition and circular-orbit domain classification for Bardeen black holes in massive gravity. Only roots with $\beta>0$ and $\Omega^2>0$ in the static exterior represent physical MSCOs.}}
\label{4}
\end{center}
\end{figure}

\begin{figure}[H]
\begin{center}
\subfigure[]{\includegraphics[width=0.98\linewidth]{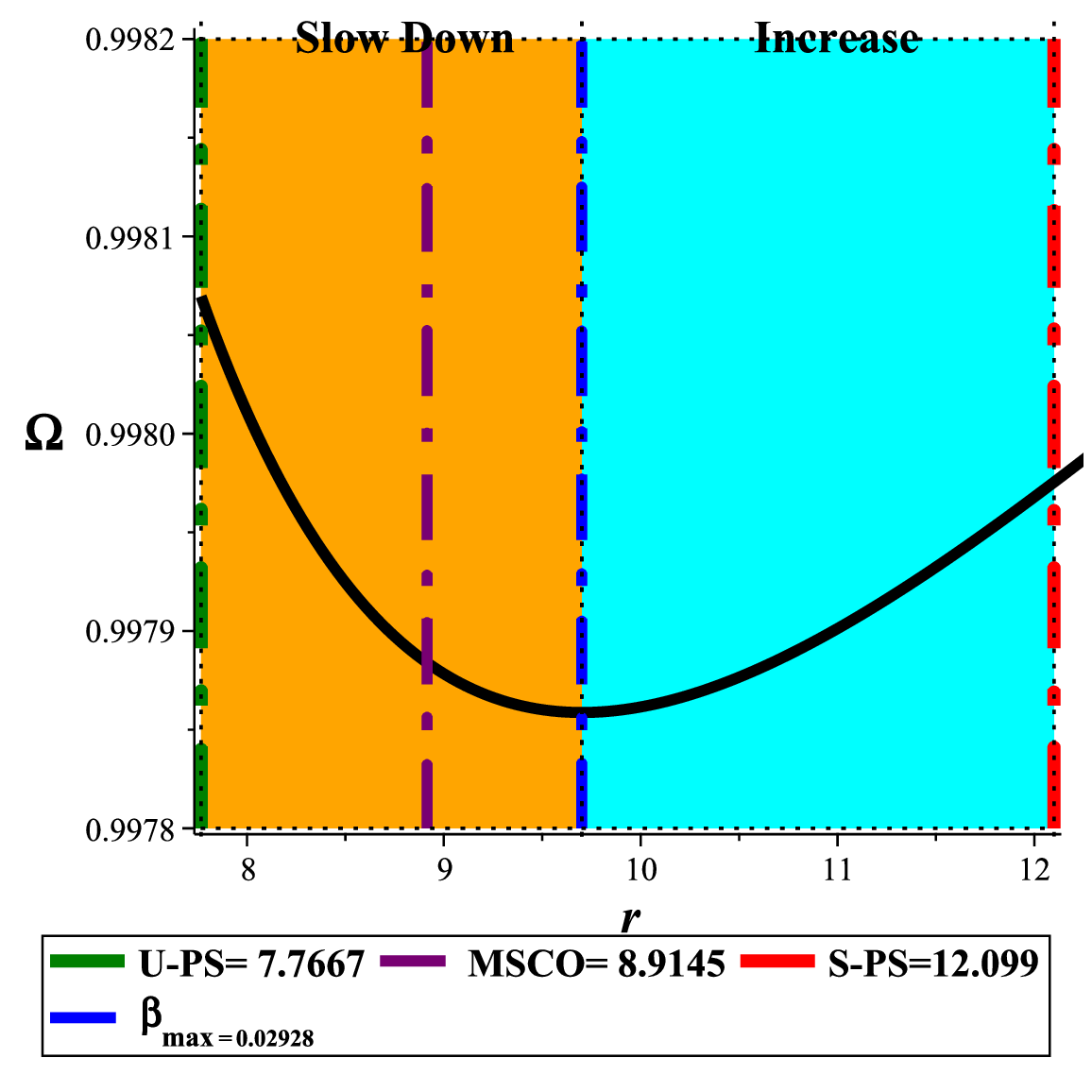}\label{5a}}
\caption{\small{Coordinate angular velocity $\Omega=d\varphi/dt$ for the Bardeen--AdS benchmark. Only the open physical TCO interval $r_{\rm U}<r<r_{\rm S}$, where $\beta>0$ and $\Omega^2>0$, is plotted; $r_{\rm U}=7.766731$ and $r_{\rm S}=12.099669$ are photon-sphere boundaries ($\beta=0$), not timelike orbits. The minimum occurs near $r=9.7006$. Parameters: $M=2, l=1, q=0.5, C=1, c_1=-1, c_2=1, m_g=0.5$.}}
\label{5}
\end{center}
\end{figure}

Figure (\ref{5}) shows that the coordinate angular velocity $\Omega(r)$ has a local minimum near $r=9.7006$ for this Bardeen--AdS parameter choice. Consistent with Eq. (\ref{eq:beta-omega-derivative}), $\beta$ has a local maximum at the same radius, with $\beta_{\max}\simeq0.02928$. The plotted curve is restricted to the physical TCO domain $7.766731<r<12.099669$, bounded by the unstable and stable photon spheres where $\beta=0$. Since $\Omega=d\varphi/dt$ is coordinate-dependent, this is a supplementary orbit property, not by itself the original Aschenbach effect. The locally measured geodesic speed is evaluated separately in Sec. \ref{subsec:local-speed-tori}.

\subsection{Non-linear Charged AdS Black Hole in Massive Gravity}

Replacing the Bardeen Lagrangian in Eq. (\ref{(12)}) with the non-linear electrodynamics Lagrangian:
\begin{equation}
\mathcal{L}(F) = F e^{-\frac{k}{2Q}\left(2Q^2F\right)^{\frac{1}{4}}},
\label{nl-Lag}
\end{equation}
yields a non-linear charged AdS black hole solution in massive gravity, with the metric function given by \cite{32}:
\begin{equation}
f(r) = 1 - \frac{2M}{r} e^{-\frac{k}{2r}} + \frac{r^2}{l^2} + m_{g}^2 \left(\frac{Cc_1 r}{2} + C^2c_2\right),
\label{(19)}
\end{equation}
where $k$ is a characteristic parameter of the non-linear electrodynamics, relating the charge $Q$ and mass $M$ via $Q^2 = Mk$. The parametric variations of the metric function are illustrated in Appendix Fig. (\ref{App-6}). While variations in $l$ and $c_1$ exert minimal influence on the horizon structure, increasing $k$ drives the system toward extremality.

\subsubsection{In Search of the Aschenbach-like Phenomenon}

To investigate the Aschenbach-like effect, we choose a baseline configuration with parameters $M = 2$, $l = 1$, $k = 0.5$, $C = 1$, $c_{1} = -1$, $c_{2} = 1$, and $m_{g} = 0.53$. Numerical evaluation of the metric function reveals two distinct horizons: a Cauchy horizon at $r = 0.0643$ and an event horizon at $r = 1.2381$. 

Using Eq. (\ref{(2)}) and Eq. (\ref{(5)}), the effective potential $H$ and the components of the vector field $\phi(r, \theta)$ are obtained as:
\begin{multline}\label{(20)}
H =\frac{1}{\sin \! \left(\theta \right) r}\Bigg[1-\frac{2 {\mathrm e}^{-\frac{k}{2 r}} M}{r}+\frac{r^{2}}{l^{2}}\\
+\left(\frac{r c_{1}}{2}+C c_{2}\right) C m_{g}^{2}\Bigg]^{1/2},
\end{multline}
\begin{multline}\label{(21)}
\phi^{r}=-\frac{\csc \! \left(\theta \right)}{r^{4}}\Bigg[\frac{M \left(k -6 r \right) {\mathrm e}^{-\frac{k}{2 r}}}{2}\\
+r^{2} \left(c_{2} m_{g}^{2} C^{2}+\frac{1}{4} r c_{1} m_{g}^{2} C +1\right)\Bigg],
\end{multline}
\begin{multline}\label{(22)}
\phi^{\Theta}=-\frac{\sqrt{2}\, \cot \! \left(\theta \right) \csc \! \left(\theta \right)}{2 r^{2}}\Bigg[\frac{-4 {\mathrm e}^{-\frac{k}{2 r}} M \,l^{2}}{r \,l^{2}}\\
+\frac{2 r \left(\left(c_{2} m_{g}^{2} C^{2}+\frac{1}{2} r c_{1} m_{g}^{2} C +1\right) l^{2}+r^{2}\right)}{r \,l^{2}}\Bigg]^{1/2}.
\end{multline}

\begin{figure}[H]
\begin{center}
\subfigure[]{\includegraphics[width=1\linewidth,height=5.7cm]{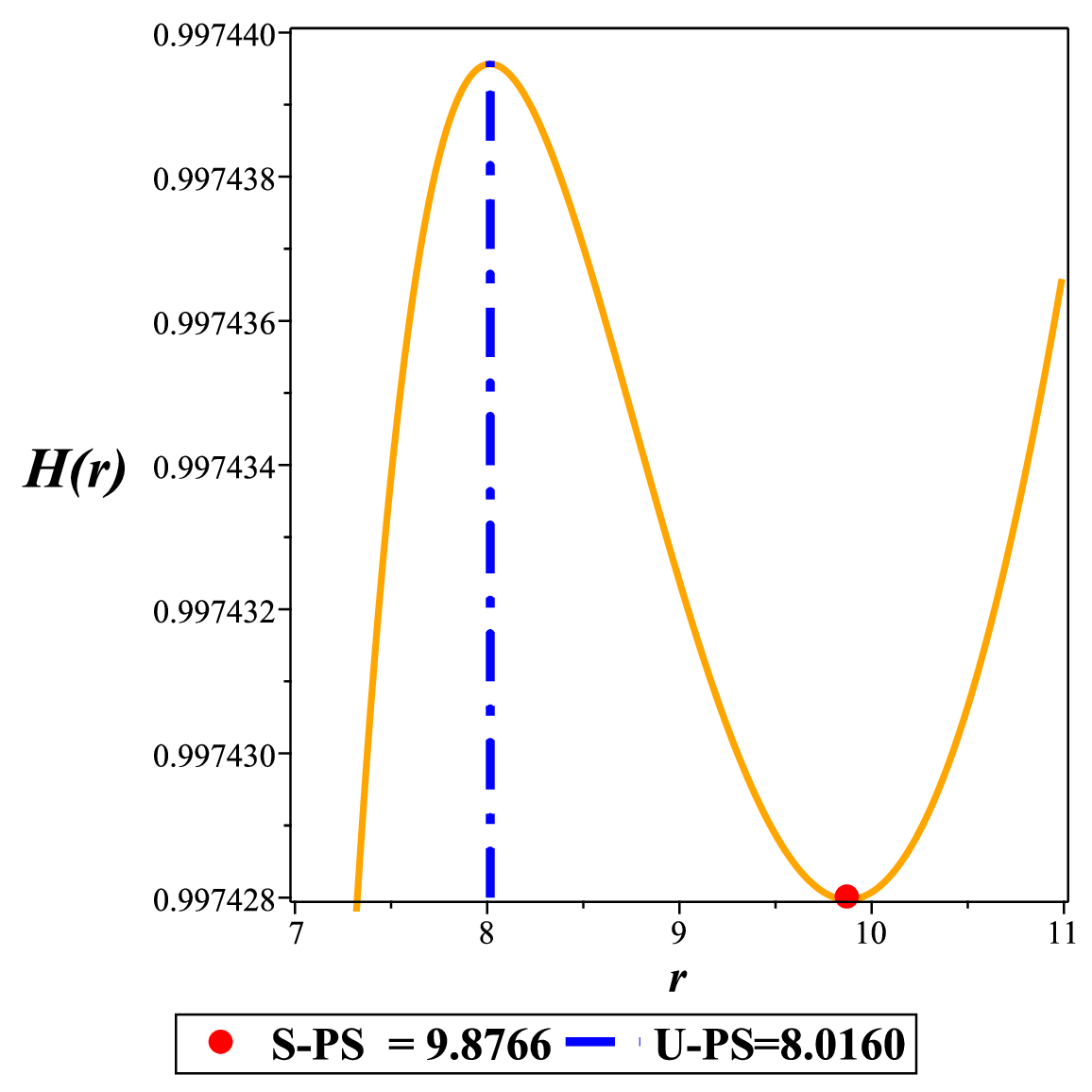}\label{7a}}\\
\subfigure[]{\includegraphics[width=1\linewidth,height=5.7cm]{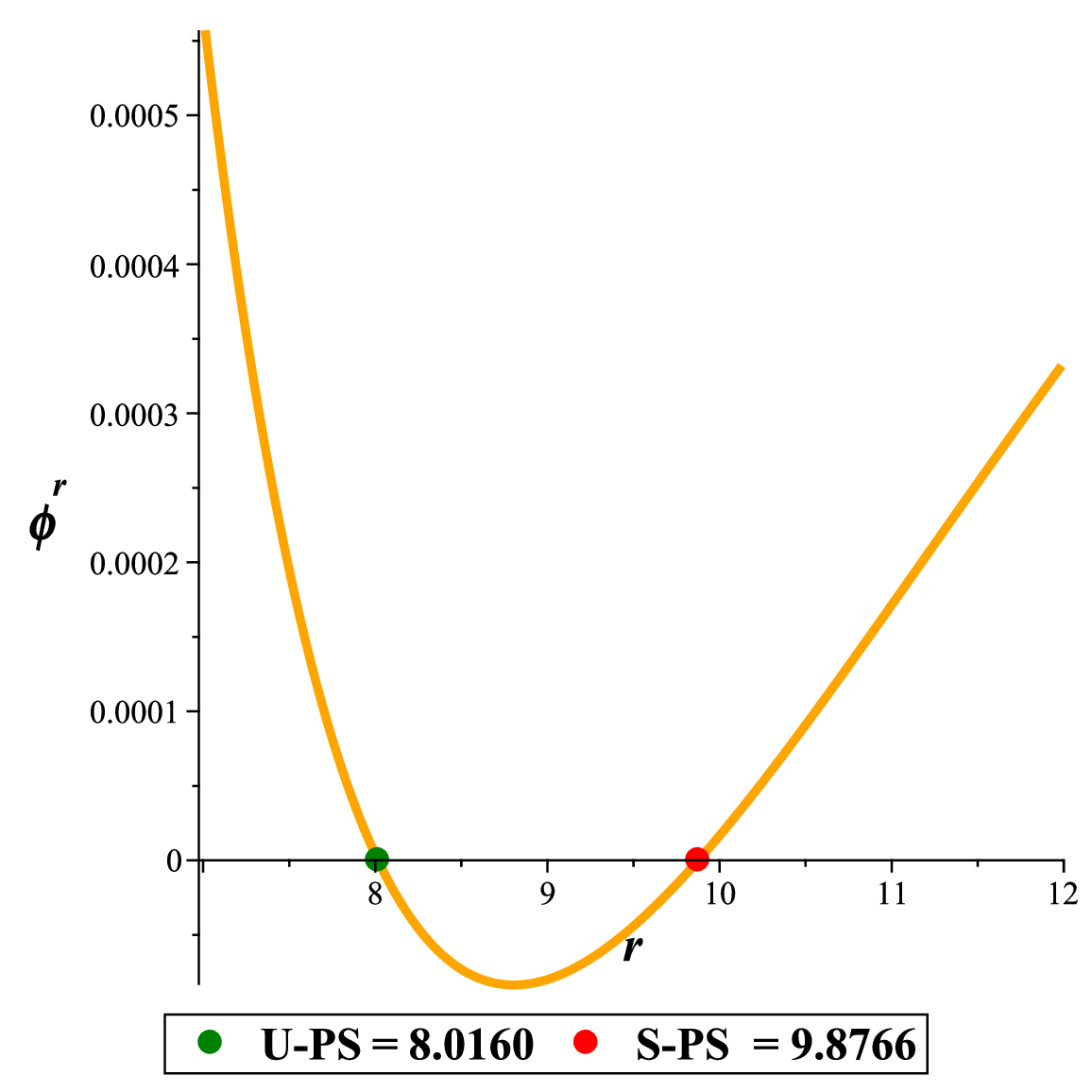}\label{7b}}
\caption{\small{(6a): Photon sphere locations at $(r, \theta) = (8.0160, 1.57)$ and $(r, \theta) = (9.8766, 1.57)$ for $M = 2, l = 1, k = 0.5, C = 1, c_{1} = -1, c_{2} = 1, m_{g} = 0.53$ in the $(r-\theta)$ plane of the normal vector field $n$; (6b): Topological potential $H(r)$ for non-linear charged AdS black holes in massive gravity.}}
\label{7}
\end{center}
\end{figure}

As demonstrated in Fig. (\ref{7}), the topological vector field indicates the presence of both the S-PS (local minimum) and the U-PS (local maximum) outside the event horizon.

\begin{figure}[H]
\begin{center}
\subfigure[]{\includegraphics[width=1\linewidth,height=6.7cm]{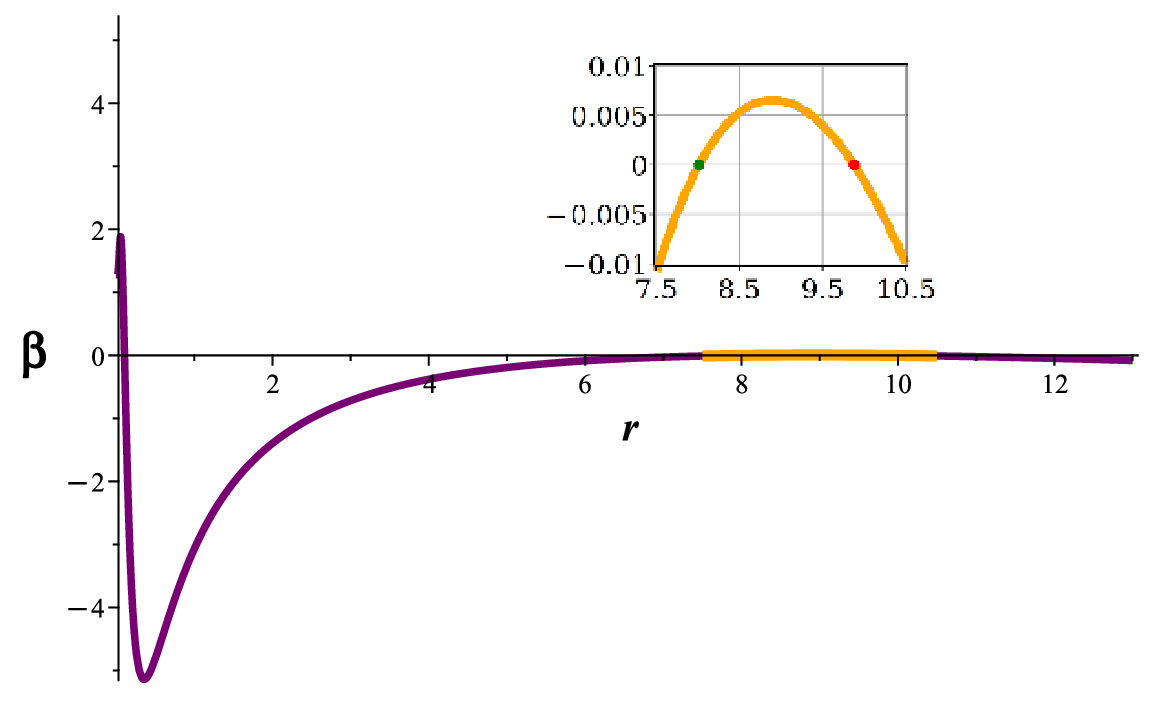}\label{8a}}\\
\subfigure[]{\includegraphics[width=1\linewidth,height=6.7cm]{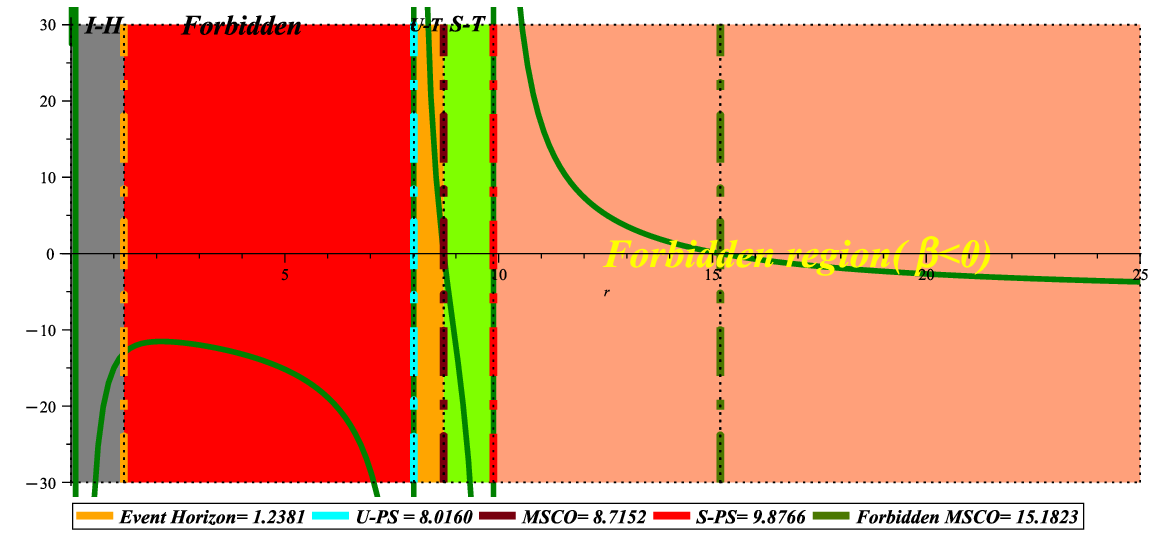}\label{8b}}
\caption{\small{(7a): $\beta$ diagram, with U-PS (blue dot) at $r = 8.0160$ and S-PS (red dot) at $r = 9.8766$; (7b): Roots of the marginal-stability condition and circular-orbit domain classification for non-linear charged AdS black holes in massive gravity. Only roots with $\beta>0$ and $\Omega^2>0$ in the static exterior represent physical MSCOs.}}
\label{8}
\end{center}
\end{figure}

The marginal-stability condition obtained from the second derivative of the effective potential $V$ has two radial roots, as shown together with $\beta$ in Fig. (\ref{8}). Only the root in the physical TCO domain is a marginally stable timelike circular orbit (MSCO). The other root lies in the $\beta<0$ region and is only a formal solution of the marginal-stability equation; it does not represent a physical timelike circular orbit.

\begin{figure}[H]
\begin{center}
\subfigure[]{\includegraphics[width=0.98\linewidth]{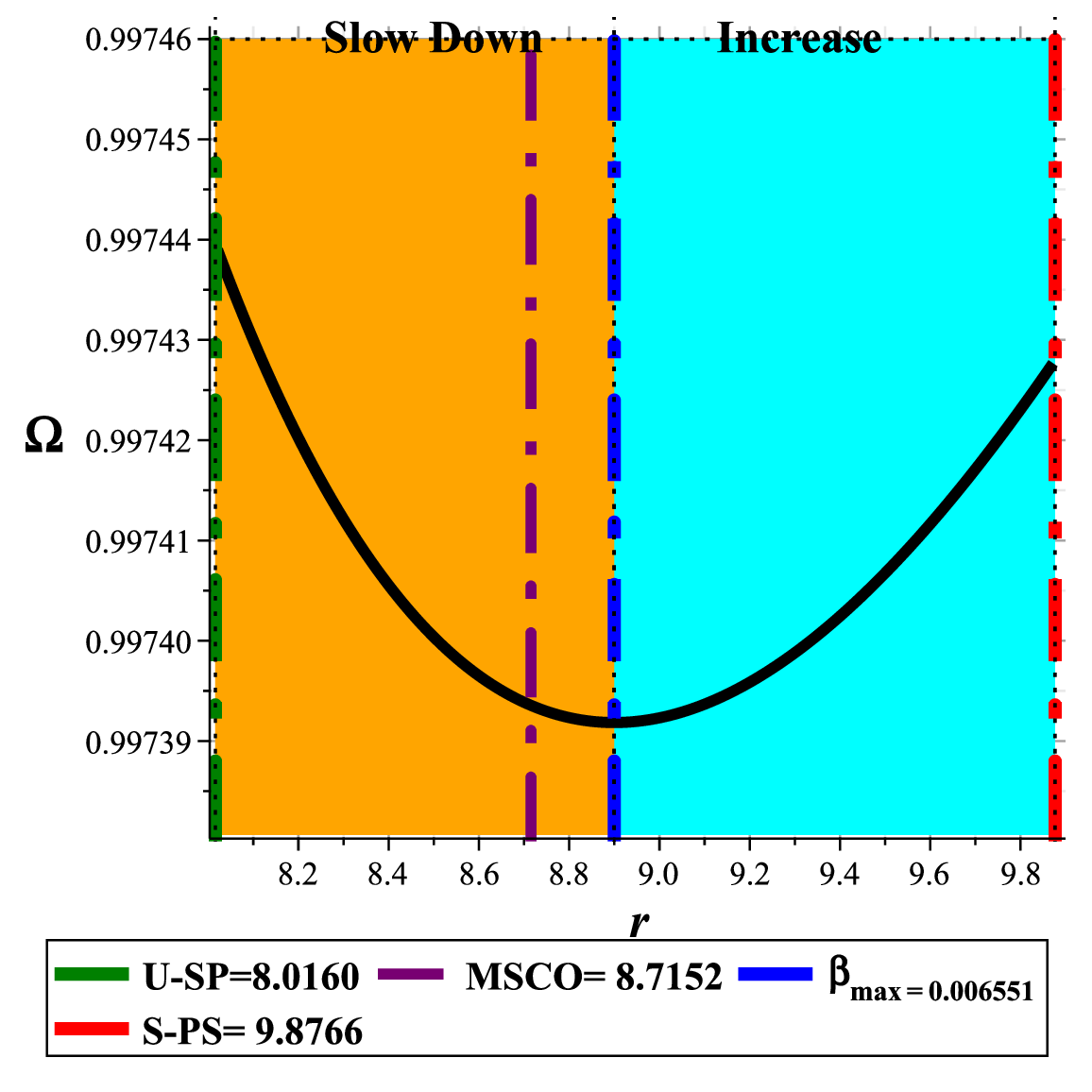}}
\caption{\small{Coordinate angular velocity $\Omega=d\varphi/dt$ for the nonlinear charged AdS benchmark. The curve is restricted to the open physical TCO interval $r_{\rm U}<r<r_{\rm S}$, where $\beta>0$ and $\Omega^2>0$; $r_{\rm U}=8.016027$ and $r_{\rm S}=9.876682$ are photon-sphere boundaries with $\beta=0$. The minimum occurs near $r=8.8998$. Parameters: $M=2, l=1, k=0.5, C=1, c_1=-1, c_2=1, m_g=0.53$.}}
\label{9}
\end{center}
\end{figure}

Figure (\ref{9}) displays a local minimum of the coordinate angular velocity near $r=8.8998$ for the chosen nonlinear charged AdS solution. Equation (\ref{eq:beta-omega-derivative}) then requires a local maximum of $\beta$ at the same radius ($\beta_{\max}\simeq0.006551$). The curve is shown only over the physical TCO domain $8.016027<r<9.876682$; the endpoint photon orbits have $\beta=0$ and are not timelike. The coordinate-dependent profile is not itself identified with the original Aschenbach effect; the locally measured speed is calculated in Sec. \ref{subsec:local-speed-tori}.

\subsection{ModMax-dRGT-like Black Hole in Massive Gravity}

Utilizing the ModMax non-linear electrodynamics Lagrangian yields a ModMax-dRGT-like black hole solution in massive gravity, with the spacetime metric given by \cite{39}:
\begin{equation}
\mathcal{L}(\mathcal{F}) = \mathcal{S}\cosh \gamma -\sqrt{\mathcal{S}^{2}+\mathcal{P}^{2}}\sinh \gamma,
\label{23}
\end{equation}
\begin{equation}
f(r) = 1 - \dfrac{M}{r} - \frac{\Lambda}{3}r^{2} + \frac{q^{2}e^{-\gamma}}{r^{2}} + m_{g}^{2}C\left( \dfrac{c_{1}r}{2}+c_{2}C\right),
\label{24}
\end{equation}
where $\mathcal{S} = \frac{\mathcal{F}}{4}$ and $\mathcal{P} = \frac{\widetilde{\mathcal{F}}}{4}$ represent the true scalar and pseudoscalar invariants, $\mathcal{F} = F_{\mu \nu }F^{\mu \nu }$ is the Maxwell invariant, $\Lambda$ is the cosmological constant, $M$ is a constant that is related to the total mass of the black hole, and $\gamma$ denotes the ModMax dimensionless parameter \cite{39}. A detailed analysis of the metric function is deferred to the Appendix, allowing us to focus directly on the physical characteristics relevant to this study.

\subsubsection{In Search of the Aschenbach-like Phenomenon}

To investigate the Aschenbach-like behavior, we choose a baseline configuration with parameters $M = 2$, $\Lambda = -0.5$, $\gamma = 2$, $q = 1$, $C = 0.4$, $c_{1} = -12$, $c_{2} = 25$, and $m_{g} = 0.5$. Numerical solution of the metric function indicates the presence of two distinct horizons: a Cauchy horizon at $r = 0.0728$ and an event horizon at $r = 1.2531$.

Using Eq. (\ref{(2)}) and Eq. (\ref{(5)}), the effective potential $H$ and the components of the vector field $\phi(r, \theta)$ are derived as:
\begin{multline}\label{(25)}
H =\frac{1}{3 \sin \! \left(\theta \right) r}\Bigg[9-\frac{9 M}{r}-3 r^{2} \Lambda +\frac{9 q^{2} {\mathrm e}^{-\gamma}}{r^{2}}\\
+9 \left(\frac{r c_{1}}{2}+C c_{2}\right) C m_{g}^{2}\Bigg]^{1/2},
\end{multline}
\begin{multline}\label{(26)}
\phi^{r}=-\frac{\csc \! \left(\theta \right)}{4 r^{4}}\Big[4 C^{2} r^{2} c_{2} m_{g}^{2}+C \,r^{3} c_{1} m_{g}^{2}\\
+8 q^{2} {\mathrm e}^{-\gamma}+4 r^{2}-6 M r \Big],
\end{multline}
\begin{multline}\label{(27)}
\phi^{\Theta}=-\frac{\sqrt{6}\, \cot \! \left(\theta \right) \csc \! \left(\theta \right)}{6 r^{2}}\Bigg[\frac{3 C \,r^{3} c_{1} m_{g}^{2}-2 r^{4} \Lambda}{r^{2}}\\
+\frac{\left(6 C^{2} c_{2} m_{g}^{2}+6\right) r^{2}-6 M r +6 q^{2} {\mathrm e}^{-\gamma}}{r^{2}}\Bigg]^{1/2}.
\end{multline}

\begin{figure}[H]
\begin{center}
\subfigure[]{\includegraphics[width=0.8\linewidth,height=8cm]{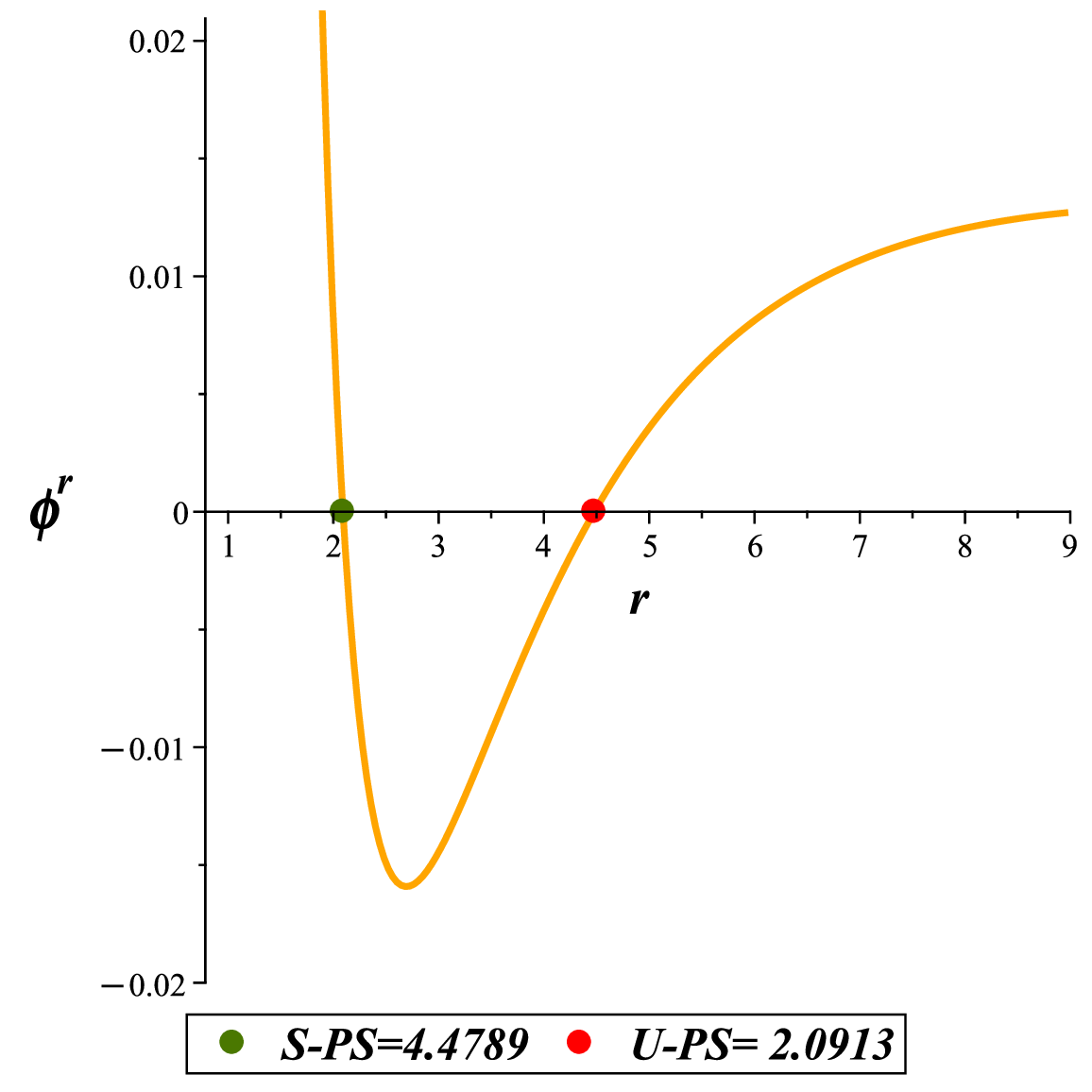}}\\
\subfigure[]{\includegraphics[width=0.8\linewidth,height=8cm]{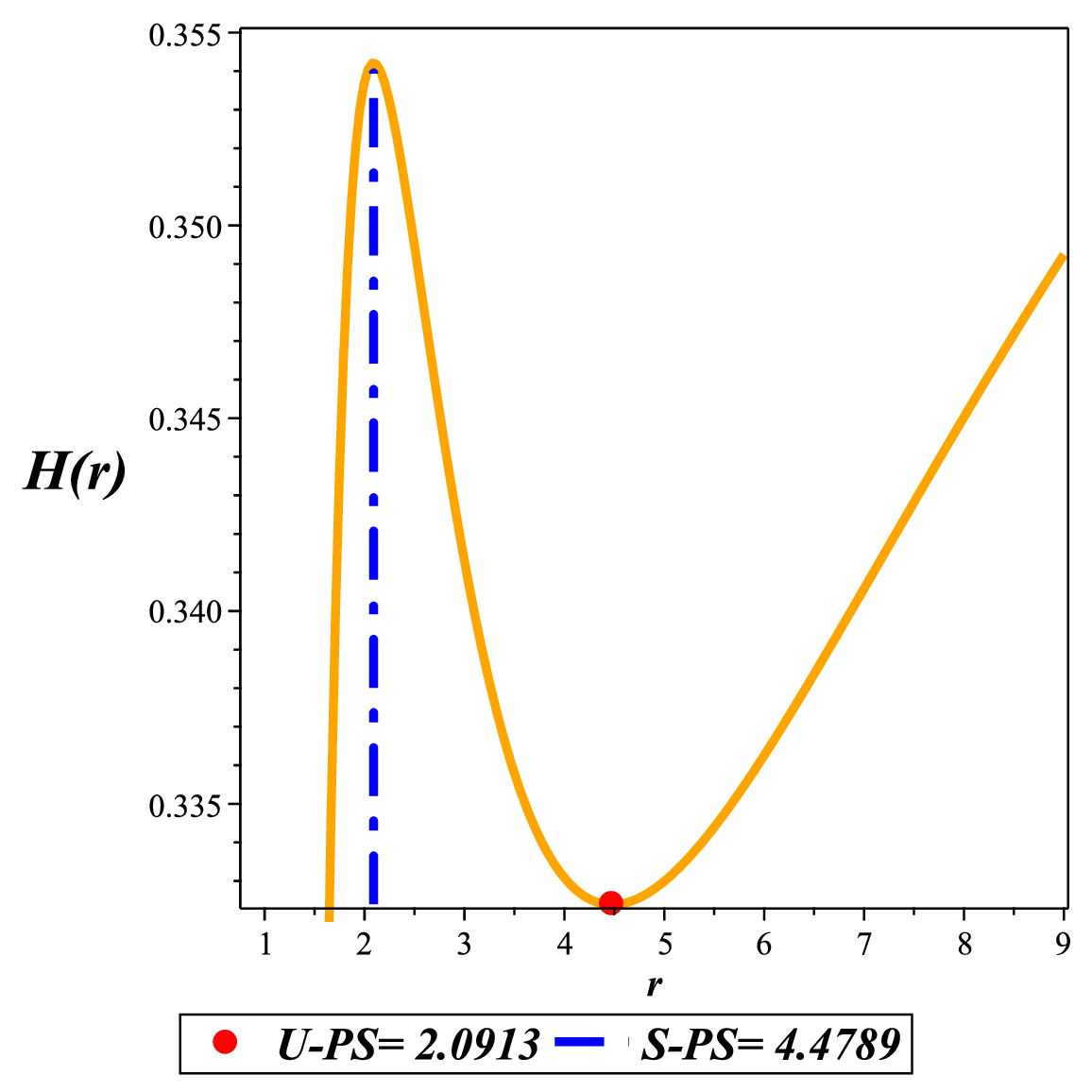}}
\caption{\small{(9a): Photon sphere locations at $(r, \theta) = (2.0913, 1.57)$ and $(r, \theta) = (4.4789, 1.57)$ for $M = 2, \Lambda = -0.5, \gamma = 2, q = 1, C = 0.4, c_{1} = -12, c_{2} = 25, m_{g} = 0.5$ in the $(r-\theta)$ plane of the normal vector field $n$; (9b): Topological potential $H(r)$ for the ModMax-dRGT-like black hole in massive gravity.}}
\label{13}
\end{center}
\end{figure}

As illustrated in Fig. (\ref{13}), the topological vector field exhibits both the S-PS (red dot) and the U-PS (green dot) outside the event horizon.

\begin{figure}[H]
\begin{center}
\subfigure[]{\includegraphics[width=1\linewidth,height=7cm]{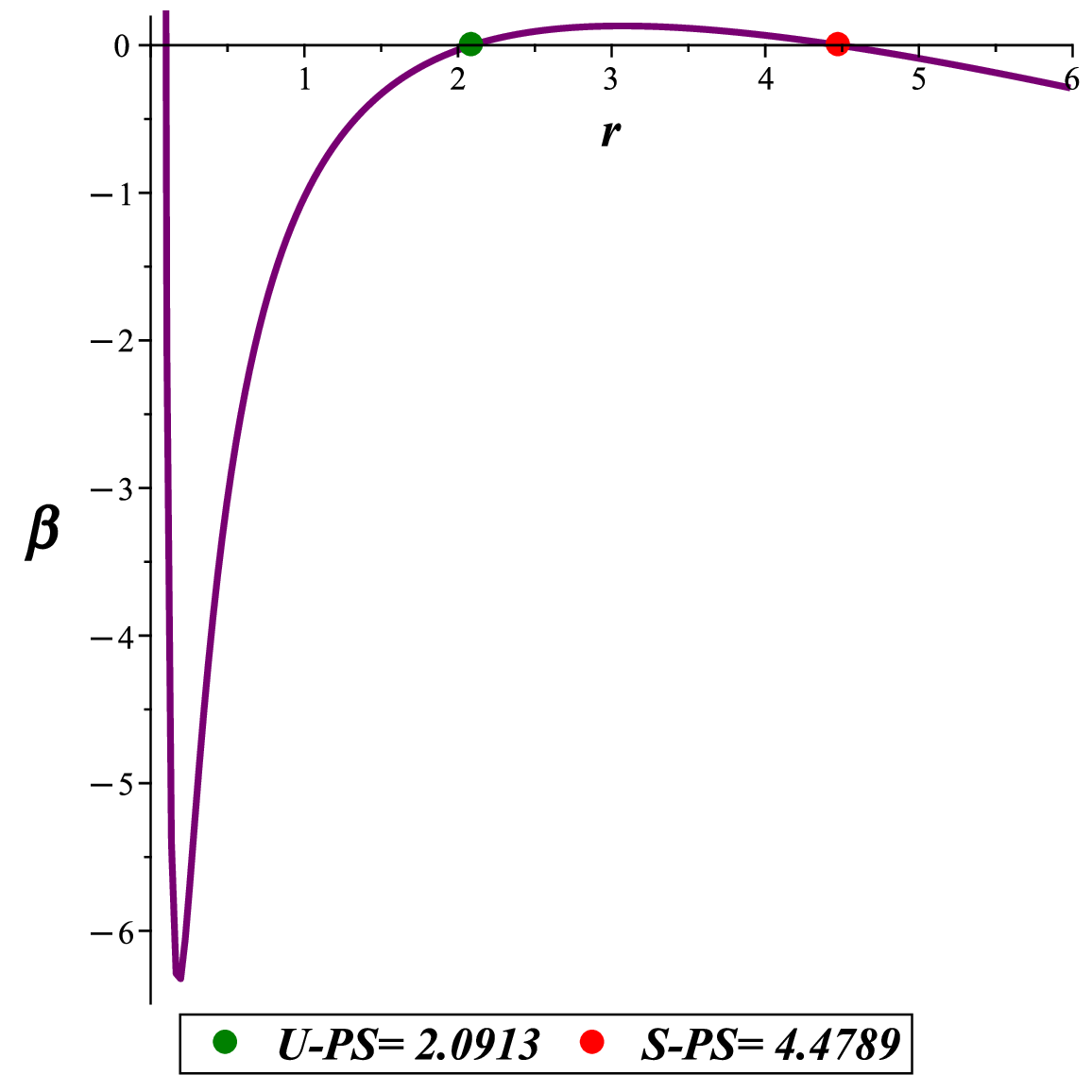}}\\
\subfigure[]{\includegraphics[width=1\linewidth,height=7cm]{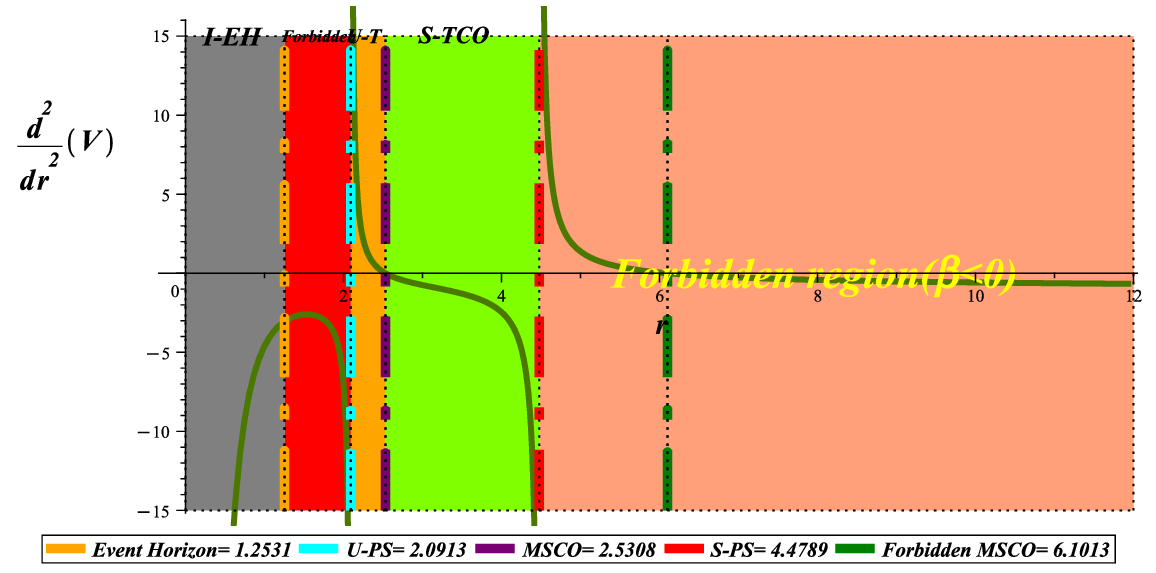}}
\caption{\small{(10a): $\beta$ diagram; (10b): Roots of the marginal-stability condition and circular-orbit domain classification for the ModMax-dRGT-like black hole in massive gravity. Only roots with $\beta>0$ and $\Omega^2>0$ in the static exterior represent physical MSCOs.}}
\label{14}
\end{center}
\end{figure}

The marginal-stability condition derived from the second derivative of the effective potential $V$ has two radial roots, as illustrated alongside $\beta$ in Fig. (\ref{14}). The root in the physical TCO domain represents an MSCO. The second root, located where $\beta<0$, is a formal root outside the admissible timelike circular-orbit domain, not a physical MSCO.

\begin{figure}[H]
\begin{center}
\subfigure[]{\includegraphics[width=0.98\linewidth]{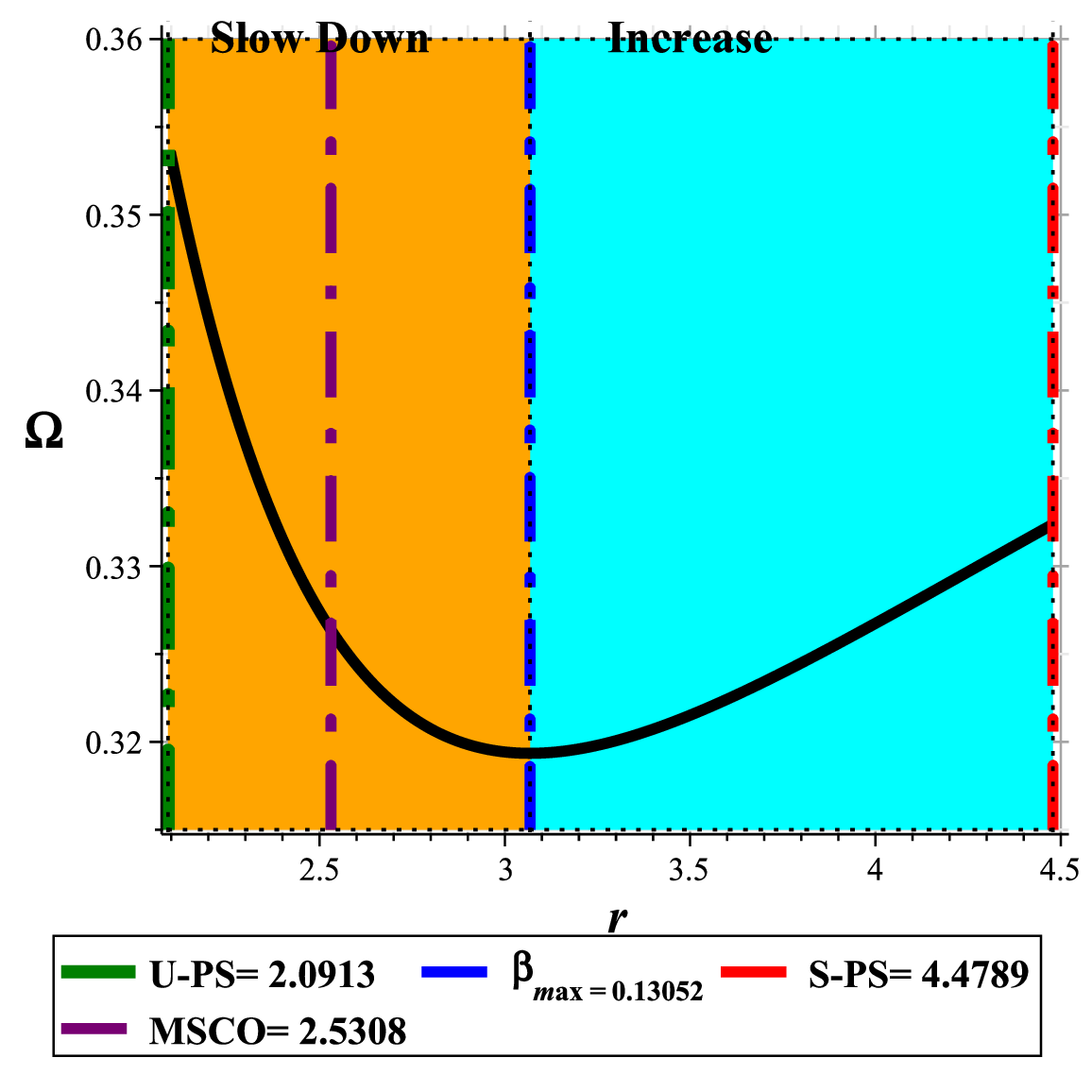}}
\caption{\small{Coordinate angular velocity $\Omega=d\varphi/dt$ for the ModMax--dRGT-like benchmark. Only the open physical TCO interval $r_{\rm U}<r<r_{\rm S}$, with $\beta>0$ and $\Omega^2>0$, is plotted; $r_{\rm U}=2.091349$ and $r_{\rm S}=4.478998$ are photon-sphere boundaries ($\beta=0$). The minimum occurs near $r=3.0679$. Parameters: $M=2, \Lambda=-0.5, \gamma=2, q=1, C=0.4, c_1=-12, c_2=25, m_g=0.5$.}}
\label{15}
\end{center}
\end{figure}

As shown in Fig. (\ref{15}), the coordinate angular velocity $\Omega(r)$ has a local minimum near $r=3.0679$ for the ModMax--dRGT-like model. Correspondingly, $\beta$ has a local maximum at the same radius ($\beta_{\max}\simeq0.13052$), as required by Eq. (\ref{eq:beta-omega-derivative}). The curve is restricted to the physical TCO domain $2.091349<r<4.478998$ between the two photon spheres; at the endpoints $\beta=0$, so they correspond to null rather than timelike circular orbits. As in the other models, this coordinate-angular-velocity profile is not identified on its own with the original Aschenbach effect; the local-speed criterion is discussed in Sec. \ref{subsec:local-speed-tori}.

\subsection{Local orbital speed and constant-specific-angular-momentum fluid tori}
\label{subsec:local-speed-tori}

\subsubsection{Locally measured speed of equatorial timelike circular geodesics}

For the static and spherically symmetric metrics considered here, the local azimuthal speed measured by a static observer is not the coordinate angular velocity $\Omega=d\varphi/dt$. On the equatorial plane, the two are related by
\begin{equation}
 v_{\rm geo}(r)=\frac{r\,\Omega(r)}{\sqrt{f(r)}}.
 \label{eq:vlocal-geodesic}
\end{equation}
For $h(r)=r^2$, the circular-orbit condition gives
\[
\Omega^2=\frac{f'(r)}{2r},
\]
and hence
\begin{equation}
 v_{\rm geo}^{\,2}(r)=\frac{r f'(r)}{2f(r)}.
 \label{eq:vgeo-f}
\end{equation}
A photon sphere satisfies $r f'(r)-2f(r)=0$, so $v_{\rm geo}=1$ at either photon-sphere radius. Between an inner unstable photon sphere at $r_{\rm U}$ and an outer stable photon sphere at $r_{\rm S}$, the effective potential $H(r,\pi/2)=\sqrt{f(r)}/r$ decreases from its local maximum to its local minimum. Therefore $r f'-2f<0$ on that interval and $v_{\rm geo}<1$. For the benchmark parameter sets used in the three model sections, direct numerical evaluation of Eq.~(\ref{eq:vgeo-f}) gives the interior minima listed in Table~\ref{tab:local-speed}. This explicitly verifies a non-monotonic locally measured speed in each example. The locations of these minima need not coincide with the minima of the coordinate angular velocity plotted in Figs.~(\ref{5}), (\ref{9}), and (\ref{15}).

\begin{table*}[t]
\centering

\caption{Comparison of the coordinate-angular-velocity minima reported in the model sections with the minima of the locally measured geodesic speed. The radii are obtained numerically from the metric functions and benchmark parameters specified in the text.}
\label{tab:local-speed}
\small
\begin{tabular}{lcccc}
\hline
Model & $r_{\Omega,\min}$ & $r_{v,\min}$ & $v_{\rm geo}(r_{v,\min})$ & $(r_{\rm U},r_{\rm S})$ \\
\hline
Bardeen--AdS & 9.700614 & 9.254047 & 0.999836 & (7.766731, 12.099669) \\
Nonlinear charged AdS & 8.899750 & 8.804245 & 0.999958 & (8.016027, 9.876682) \\
ModMax--dRGT-like & 3.067868 & 2.732506 & 0.931076 & (2.091349, 4.478998) \\
\hline
\end{tabular}
\end{table*}

\subsubsection{Constant-specific-angular-momentum fluid tori}

Following the analysis of constant-specific-angular-momentum fluid tori in Ref.~\cite{43}, we examine non-geodesic circular motion of a perfect fluid with constant specific angular momentum. Consider a stationary, circular flow with four-velocity $u^\mu=u^t(1,0,0,\Omega)$ and constant
\begin{equation}
 \ell_0\equiv-\frac{u_\varphi}{u_t}=\mathrm{const.}
\end{equation}
For the metrics in this paper, $g_{t\varphi}=0$ and $g_{\varphi\varphi}=r^2\sin^2\theta$, which imply
\begin{equation}
 \Omega=\frac{\ell_0 f(r)}{r^2\sin^2\theta}.
\end{equation}
The azimuthal speed measured by a static local observer is therefore
\begin{equation}
 v_{\rm fluid}^{(\varphi)}=\frac{\sqrt{g_{\varphi\varphi}}\,\Omega}{\sqrt{-g_{tt}}}
 =\ell_0\frac{\sqrt{f(r)}}{r\sin\theta}
 =\ell_0 H(r,\theta).
 \label{eq:vfluid-H}
\end{equation}
Thus, at fixed polar angle and for $\ell_0>0$, the stationary points of the local fluid-speed profile coincide with those of $H$. In particular, in the equatorial plane,
\begin{equation}
 \frac{d v_{\rm fluid}^{(\varphi)}}{dr}
 =\frac{\ell_0\,[r f'(r)-2f(r)]}{2r^2\sqrt{f(r)}}.
 \label{eq:vfluid-derivative}
\end{equation}
The speed is physically admissible only where $f(r)>0$ and $1-\ell_0^2H^2(r,\theta)>0$, equivalently $|v_{\rm fluid}^{(\varphi)}|<1$.

To check that the local-speed extremum can occur inside a fluid configuration rather than only in the vacuum geometry, we use the standard relativistic potential for a barotropic perfect fluid with constant specific angular momentum,
\begin{equation}
 W(r,\theta)=\ln|u_t|
 =\frac12\ln\!\left[\frac{f(r)}{1-\ell_0^2 H^2(r,\theta)}\right].
 \label{eq:fluid-potential}
\end{equation}
On the equatorial plane, its critical points satisfy
\begin{equation}
 \ell_0^2=\ell_K^2(r)\equiv\frac{r^3 f'(r)}{2f^2(r)},
 \label{eq:ellK}
\end{equation}
where $\ell_K(r)$ is the specific angular momentum of a circular geodesic.

For each benchmark model, we define
\[
\ell_{\min}=\min_{r_{\rm U}<r<r_{\rm S}}\ell_K(r),
\qquad
\ell_{\lim}=\frac{1}{H(r_{\rm U})}=\ell_K(r_{\rm U}),
\]
and choose
\[
\ell_0=\ell_{\min}+0.85(\ell_{\lim}-\ell_{\min})
\]
as an illustrative value. The coefficient $0.85$ is not a physically preferred value. It is chosen because, for all three benchmark metrics, it yields the ordering
\[
W_{\rm eq}(r_{\rm center})<W_{\rm eq}(r_{\rm S})
<W_{\rm in}<W_{\rm eq}(r_{\rm cusp}),
\]
so that the selected equipotential contains the stable photon-sphere radius while remaining below the inner potential barrier. Other values may also be admissible, but their suitability must be checked separately for each metric.

For this choice, the equatorial potential $W_{\rm eq}(r)=W(r,\pi/2)$ has an inner local maximum at $r_{\rm cusp}$ and an outer local minimum at $r_{\rm center}$. The point $r_{\rm cusp}$ is the cusp of the critical equipotential; the closed equipotential selected below lies below this critical level and therefore does not itself possess a cusp. We choose the equatorial surface level
\[
W_{\rm in}=\frac{W_{\rm eq}(r_{\rm cusp})+W_{\rm eq}(r_{\rm S})}{2},
\]
and solve $W_{\rm eq}(r)=W_{\rm in}$ for the inner and outer equatorial edges, $r_{\rm in}$ and $r_{\rm out}$. The resulting torus parameters and the maximum local speeds at the equatorial edges for the three benchmark models are summarized in Table~\ref{tab:fluid-tori}.

\begin{table*}[t]
\centering

\caption{Illustrative constant-$\ell_0$ equilibrium-torus configurations for the three benchmark geometries. Here $r_{\rm cusp}$ and $r_{\rm center}$ are the inner local maximum and outer local minimum of the equatorial potential $W_{\rm eq}(r)$, and $r_{\rm in}$ and $r_{\rm out}$ are the equatorial intersections of the selected equipotential. The last column gives the maximum local speed at the two equatorial edges.}
\label{tab:fluid-tori}
\scriptsize
\resizebox{\textwidth}{!}{%
\begin{tabular}{lcccccccc}
\hline
Model & $\ell_{\min}$ & $\ell_{\lim}$ & $\ell_0$ & $r_{\rm cusp}$ & $r_{\rm center}$ & $r_{\rm in}$ & $r_{\rm out}$ & $\max |v_{\rm fluid}^{(\varphi)}|$ \\
\hline
Bardeen--AdS & 1.001803415 & 1.001934189 & 1.001914573 & 7.839994 & 10.729794 & 8.621072 & 13.884556 & 0.999963 \\
Nonlinear charged AdS & 1.002529829 & 1.002567008 & 1.002561431 & 8.063918 & 9.602866 & 8.615438 & 10.578173 & 0.999991 \\
ModMax--dRGT-like & 2.680430021 & 2.823134190 & 2.801728564 & 2.118344 & 3.272638 & 2.343892 & 5.289269 & 0.986171 \\
\hline
\end{tabular}%
}
\end{table*}

For all three numerical examples, the calculated potential levels obey
\begin{equation}
 W_{\rm eq}(r_{\rm center})
 <
 W_{\rm eq}(r_{\rm S})
 <
 W_{\rm in}
 <
 W_{\rm eq}(r_{\rm cusp}).
\end{equation}
The entire selected equatorial interval satisfies $f(r)>0$ and $\ell_0^2H^2(r,\pi/2)<1$. Thus the stable photon-sphere radius lies inside the equatorial section of the torus, the selected equipotential remains below the inner critical-potential barrier, and the fluid velocity remains subluminal throughout the displayed equatorial interval. Since $v_{\rm fluid}^{(\varphi)}=\ell_0 H$, it decreases toward its local minimum at $r_{\rm S}$ and increases beyond it. Hence the non-monotonic profile is present within the fluid torus itself.

The Bardeen--AdS and nonlinear charged AdS examples have edge speeds very close to the speed of light because the illustrative values of $\ell_0$ lie close to the limiting value. This demonstrates mathematical admissibility, not astrophysical realism or stability against hydrodynamic perturbations. We have not analyzed perturbative stability, a specific equation of state, radiation transfer, or the excitation of QPO modes. Such an analysis would be needed before linking these tori to a specific observed signal.

The closed geometry of the selected equipotential follows from its meridional level-set equation.  Define
\[
\mathcal A(r)\equiv\frac{\ell_0^2f(r)}{r^2}.
\]
For each benchmark, the numerical equatorial potential decreases from $r_{\rm in}$ to $r_{\rm center}$ and increases from $r_{\rm center}$ to $r_{\rm out}$, with
\begin{equation}
\begin{gathered}
 W_{\rm eq}(r_{\rm in})=W_{\rm eq}(r_{\rm out})=W_{\rm in},\\
 W_{\rm eq}(r)<W_{\rm in}
 \qquad (r_{\rm in}<r<r_{\rm out}),\\
 f(r)>0,
 \qquad
 0<\mathcal A(r)<1.
\end{gathered}
\label{eq:torus-equatorial-closure}
\end{equation}
Solving Eq.~(\ref{eq:fluid-potential}) at $W=W_{\rm in}$ gives
\begin{equation}
 Q(r)\equiv\sin^2\theta_{\rm surf}(r)
 =
 \frac{\mathcal A(r)}
 {1-f(r)e^{-2W_{\rm in}}}.
\label{eq:torus-surface-closure}
\end{equation}
For $r_{\rm in}<r<r_{\rm out}$, the condition $W_{\rm eq}(r)<W_{\rm in}$ implies
\[
f(r)e^{-2W_{\rm in}}<1-\mathcal A(r).
\]
Since $\mathcal A(r)>0$, the denominator in Eq.~(\ref{eq:torus-surface-closure}) is positive and exceeds the numerator. Consequently,
\[
0<Q(r)<1
\qquad
(r_{\rm in}<r<r_{\rm out}),
\]
while
\[
Q(r_{\rm in})=Q(r_{\rm out})=1.
\]

The two continuous branches
\[
\theta_-(r)=\arcsin\sqrt{Q(r)},
\qquad
\theta_+(r)=\pi-\theta_-(r)
\]
therefore join at $r_{\rm in}$ and $r_{\rm out}$ on the equatorial plane, forming a closed meridional curve surrounding $(r_{\rm center},\pi/2)$. At both endpoints, the equatorial potential level is crossed transversely, since
\[
W_{\rm eq}'(r_{\rm in})\ne0,
\qquad
W_{\rm eq}'(r_{\rm out})\ne0.
\]
Thus the two branches form a regular closed meridional contour, with the expected vertical tangents in the representation $\theta_\pm(r)$ at the equatorial endpoints. Since $r_{\rm in}>r_h$ and $Q(r)>0$ throughout the closed interval, the contour touches neither the horizon nor the symmetry axis. Its azimuthal rotation therefore generates a closed toroidal surface. This conclusion concerns the connected component surrounding the chosen centre, not any disconnected components of the same level set.

The entire enclosed component is also kinematically admissible. Between the two surface branches, $\sin^2\theta\ge Q(r)$, and hence
\begin{equation}
 \bigl(v_{\rm fluid}^{(\varphi)}\bigr)^2
 =
 \frac{\mathcal A(r)}{\sin^2\theta}
 \le
 1-f(r)e^{-2W_{\rm in}}
 <1.
\label{eq:torus-full-subluminality}
\end{equation}
Thus the subluminality check extends to the full torus, rather than only to its equatorial section. Closure and timelike motion do not imply hydrodynamic or spacetime perturbative stability.

\section{Conclusions}

The results establish that the three benchmark static black-hole geometries considered here can support both non-monotonic coordinate-angular-velocity profiles for circular geodesics and non-monotonic locally measured azimuthal-speed profiles. The latter is the appropriate kinematic quantity for comparison with the original Aschenbach effect. The distinction matters because the coordinate angular velocity of co-rotating equatorial circular geodesics in Kerr spacetime is monotonic, whereas the original Aschenbach reversal concerns the velocity measured in the LNRF.

We also found, within the assumptions of a barotropic perfect fluid in stationary circular motion with constant specific angular momentum, illustrative equilibrium tori that contain the stable photon-sphere radius and exhibit a non-monotonic local azimuthal-speed profile. These calculations demonstrate a possible fluid analogue at the level of equilibrium configurations; they do not establish the dynamical stability of the tori or the existence of a particular oscillation mode.

Although QPO models based on characteristic frequencies of flares from the Galactic Centre and on the LNRF-velocity hump in nearly extreme Kerr spacetime have been discussed in the literature \cite{40,41,42}, the present work does not calculate a corresponding set of mode frequencies, compare them with observed QPO frequencies, or model the emitted radiation. It therefore does not claim to explain the QPOs observed in Sgr A* or microquasars. The non-monotonic profiles reported here provide motivation for a future quantitative study of fluid perturbations and frequency scaling, but any observational application remains open. The conclusions are restricted to the static metrics and parameter choices analyzed.

A velocity extremum is not itself an oscillation frequency. Applying these examples to HF QPOs would require a specified fluid-perturbation model, identification of the excited modes, and calculation of their frequencies with respect to a stated time coordinate. Those frequencies would then have to be converted into physical units using a justified mass and parameter scaling, with the observer-dependent redshift and emitted signal modeled before comparison with data. In addition, the benchmark metrics have AdS-type asymptotics: the first two use a finite AdS radius, and the third uses $\Lambda<0$. A direct identification with a microquasar or Sgr A* therefore requires a justified astrophysical interpretation of these backgrounds and their time normalization; the present examples alone do not provide that identification. The Kerr proposals in Refs.~\cite{40,41,42} cannot be transferred to these static models merely because both exhibit a non-monotonic local-speed profile.

A rotating extension is not obtained by simply adding a spin parameter to the static metrics used here. It would require a consistent stationary, axisymmetric solution of the relevant field equations, followed by a new analysis of null circular orbits, timelike circular motion, and the appropriate locally measured velocity. In particular, the existence and stability of the photon spheres found in the static solutions need not persist in a rotating geometry. We therefore leave this distinct problem to a separate investigation rather than infer a rotating result from the present static analysis.

\section{Appendix}

\begin{figure}[H]
\begin{center}
\subfigure[]{\includegraphics[width=0.9\linewidth,height=3.7cm]{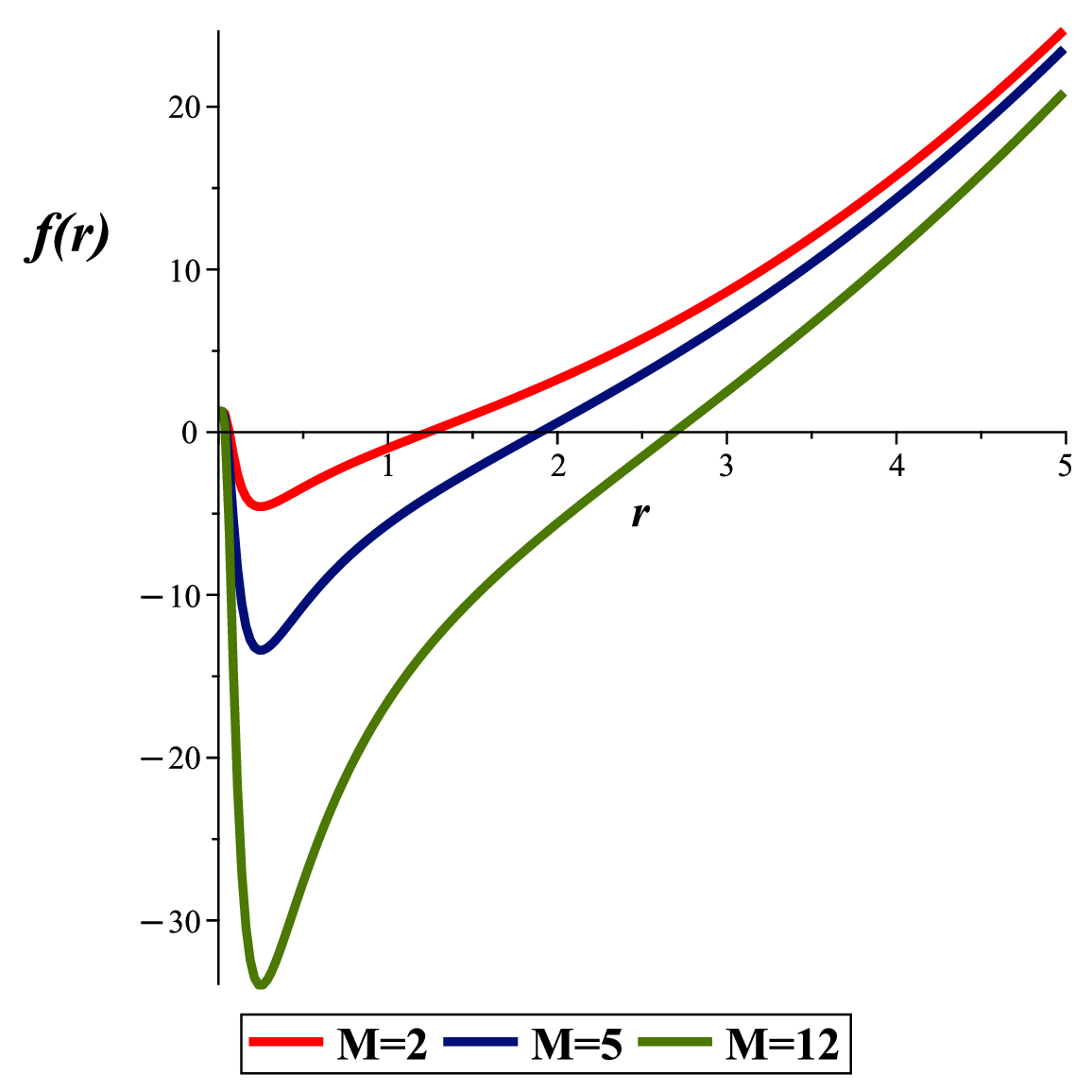}\label{6a}}\\
\subfigure[]{\includegraphics[width=0.9\linewidth,height=3.7cm]{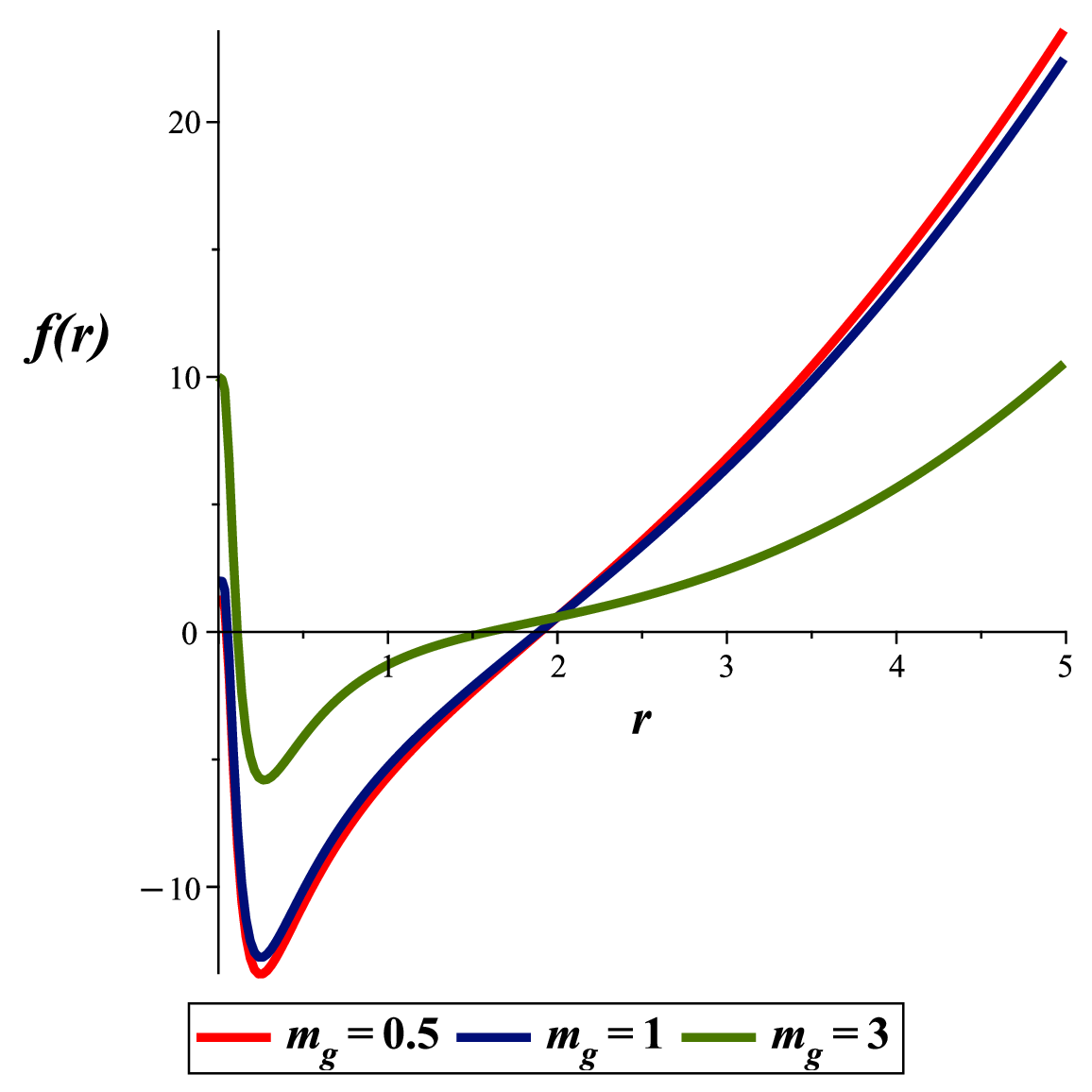}\label{6b}}\\
\subfigure[]{\includegraphics[width=0.9\linewidth,height=3.7cm]{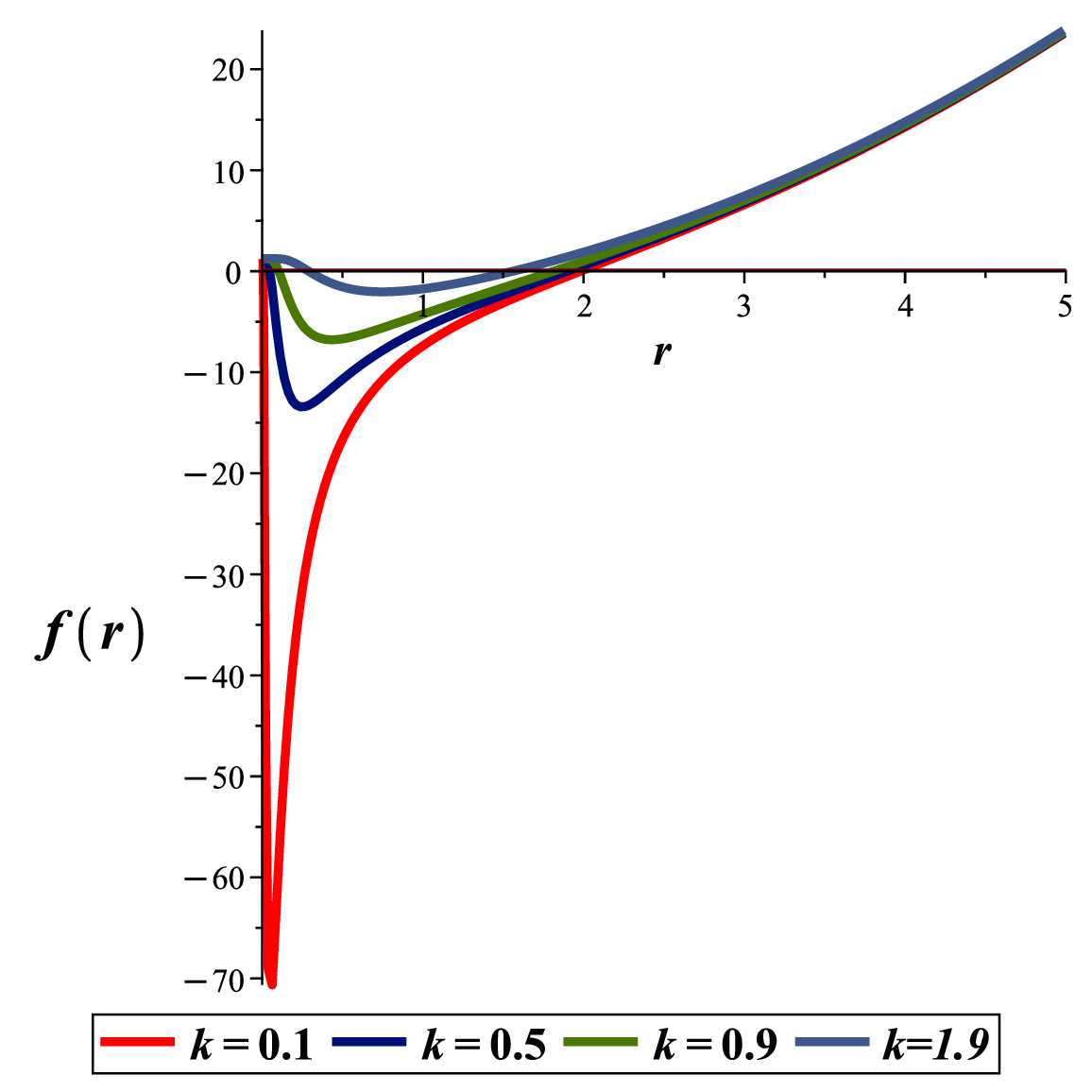}\label{6c}}\\
\subfigure[]{\includegraphics[width=0.9\linewidth,height=3.7cm]{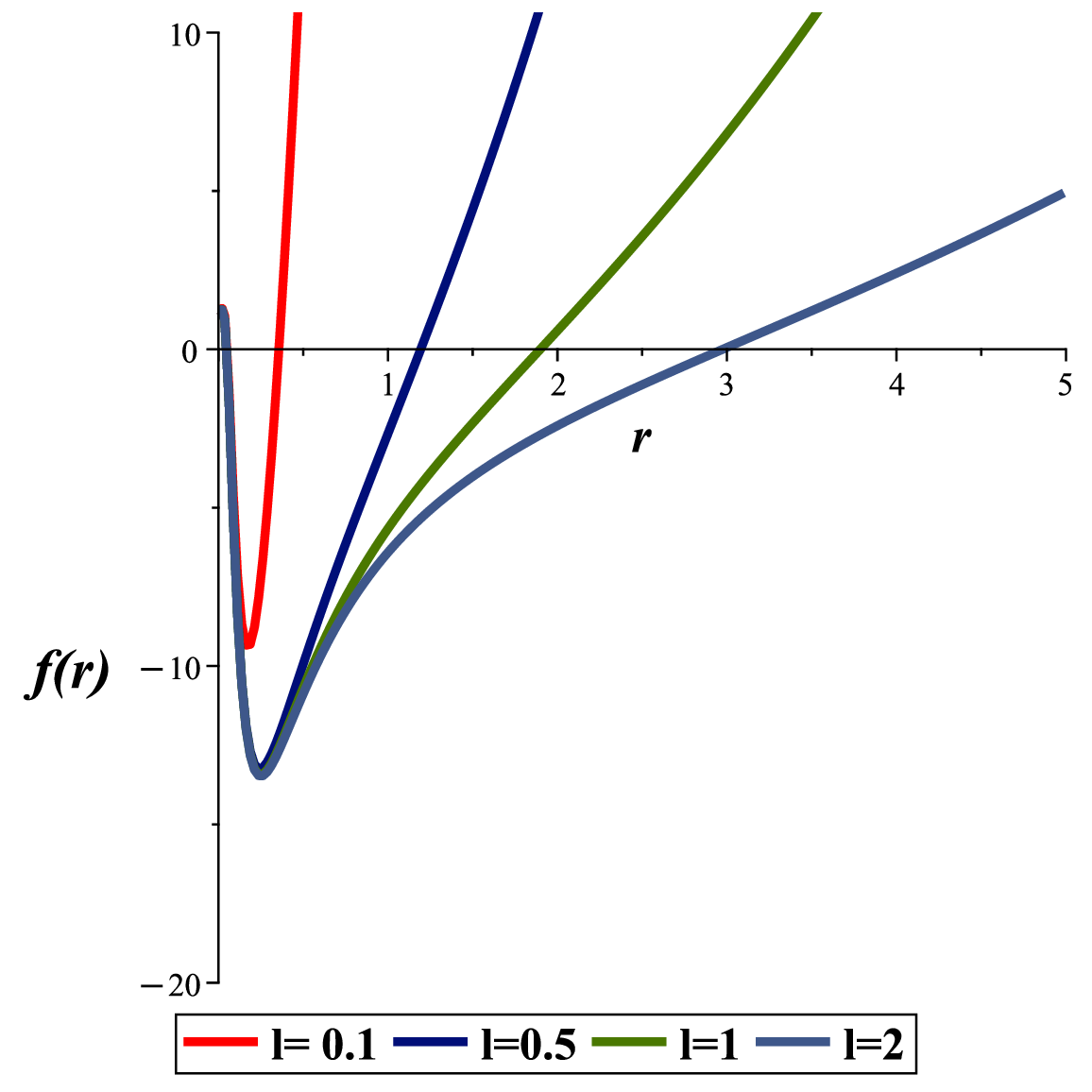}\label{6d}}\\
\subfigure[]{\includegraphics[width=0.9\linewidth,height=3.7cm]{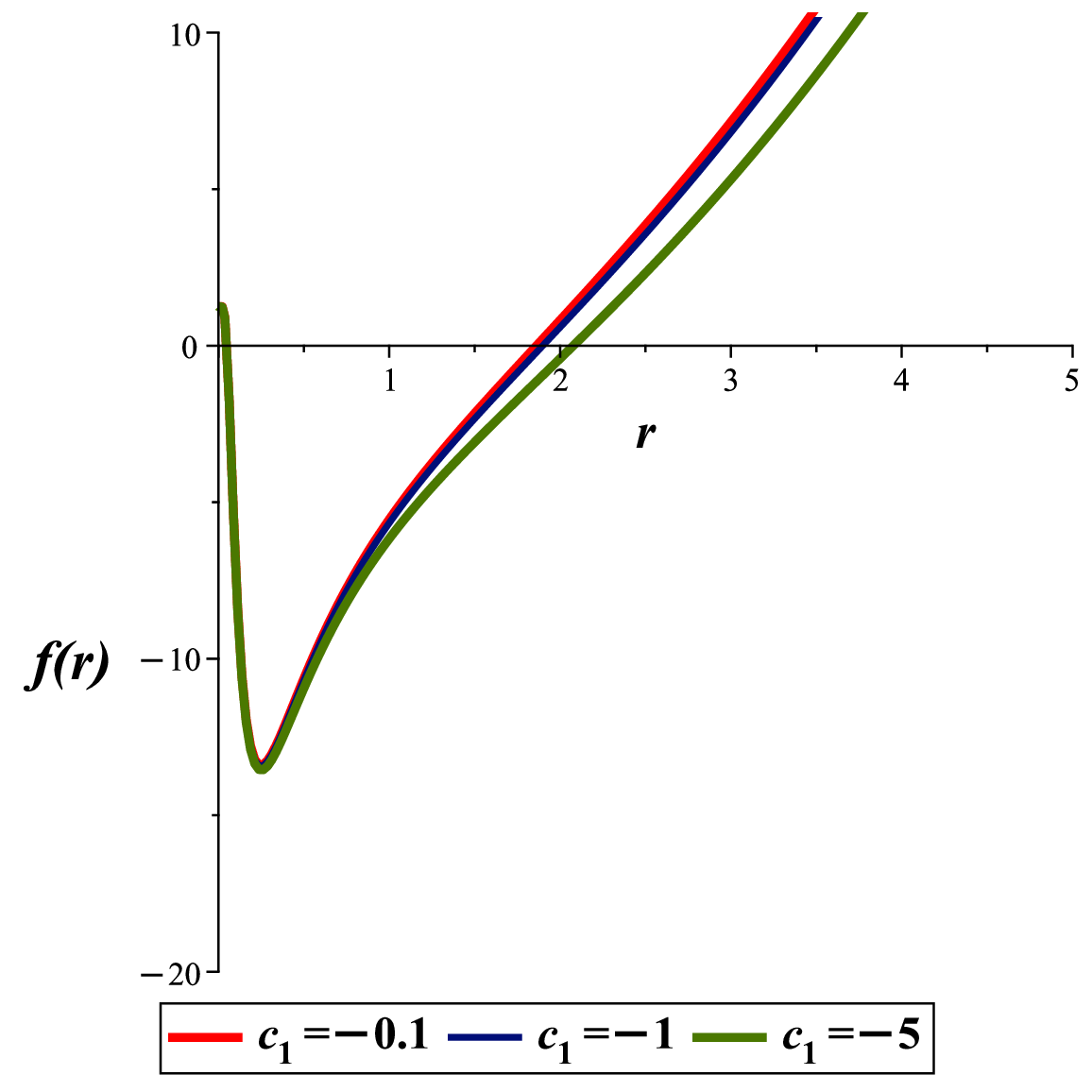}\label{6e}}
\caption{\small{Metric function with $l = 1, k = 0.5, C = 1, c_{1} = -1, c_{2} = 1$: (a) with $m_{g} = 0.53$ and varying $M$; (b) with $M = 5$ and varying $m_{g}$. For $M = 5, C = 1, c_{1} = -1, c_{2} = 1, m_{g} = 0.5$: (c) with $l = 1$ and varying $k$; (d) with $k = 0.5$ and varying $l$; and (e) with $l = 1, k = 0.5$ and varying $c_{1}$ for non-linear charged AdS black holes in massive gravity.}}
\label{App-6}
\end{center}
\end{figure}

\begin{center}
\textbf{Non-linear Charged AdS Black Hole in Massive Gravity}
\end{center}
The behavior of the metric function for the non-linear charged AdS black hole in massive gravity under various parametric variations is illustrated in Fig. (\ref{App-6}).

\begin{figure}[H]
\begin{center}
\subfigure[]{\includegraphics[width=0.8\linewidth,height=8cm]{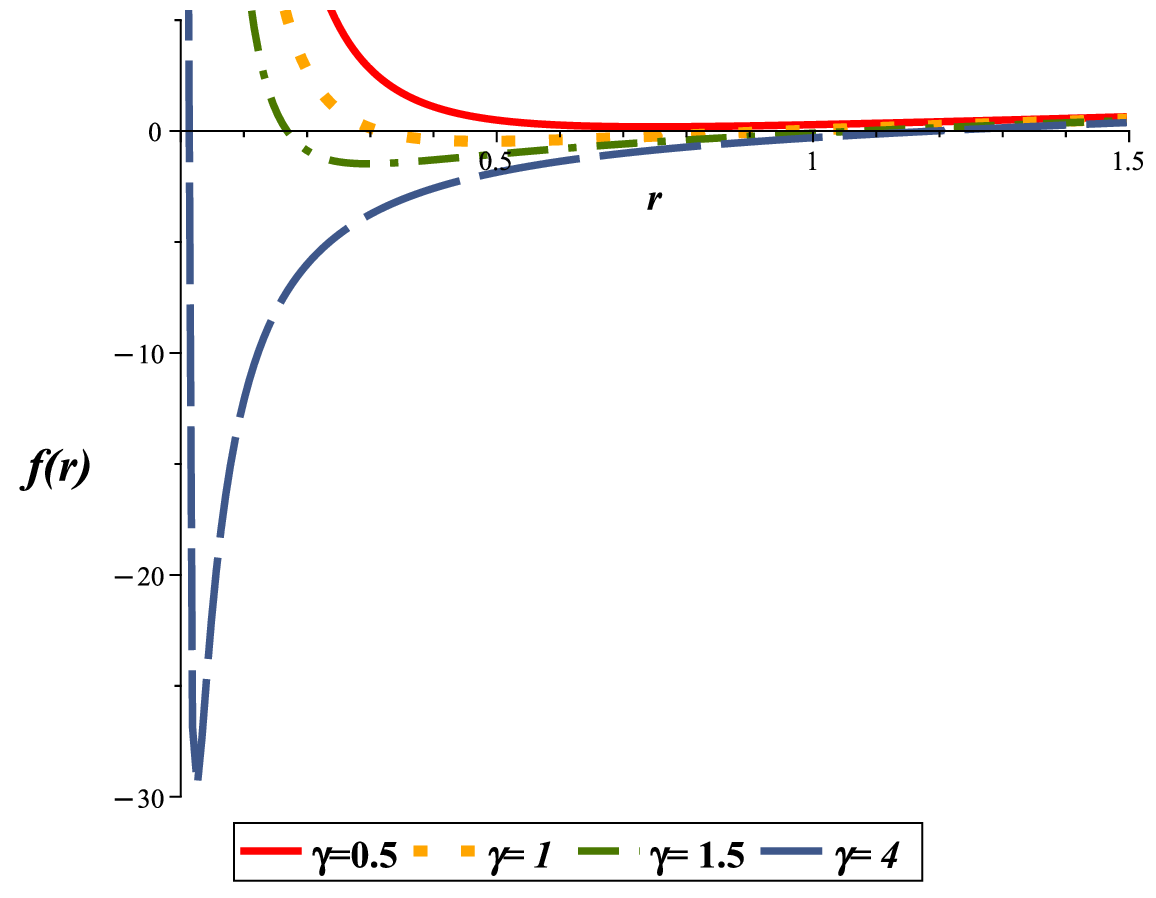}\label{10a}}
\caption{\small{Metric function with $M = 1.5, \Lambda = -0.5, q = 1, C = 0.4, c_{1} = -12, c_{2} = 25, m_{g} = 0.1$ for different values of $\gamma$ in the ModMax-dRGT-like black hole in massive gravity.}}
\label{App-10}
\end{center}
\end{figure}

\begin{center}
\textbf{ModMax-dRGT-like Black Hole in Massive Gravity}
\end{center}
As shown in Fig. (\ref{App-10}), the parameter $\gamma$ significantly influences the gravitational characteristics of the model. Small values ($\gamma < 0.6686$) drive the system toward extremality, whereas larger values promote black hole behavior. Consequently, the metric function is analyzed in two distinct regimes.

\begin{figure}[H]
\begin{center}
\subfigure[]{\includegraphics[width=1\linewidth,height=3.7cm]{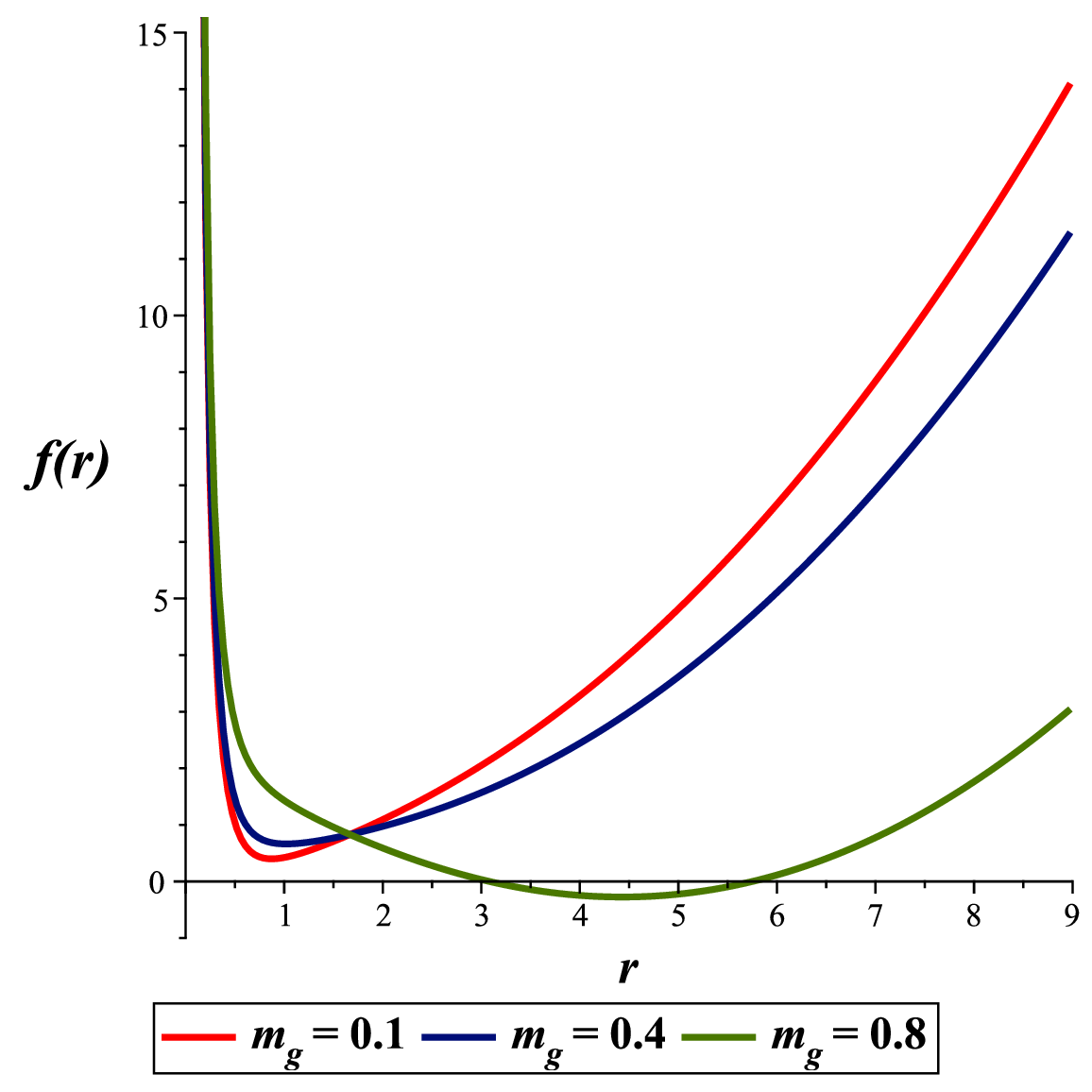}\label{11a}}\\
\subfigure[]{\includegraphics[width=1\linewidth,height=3.7cm]{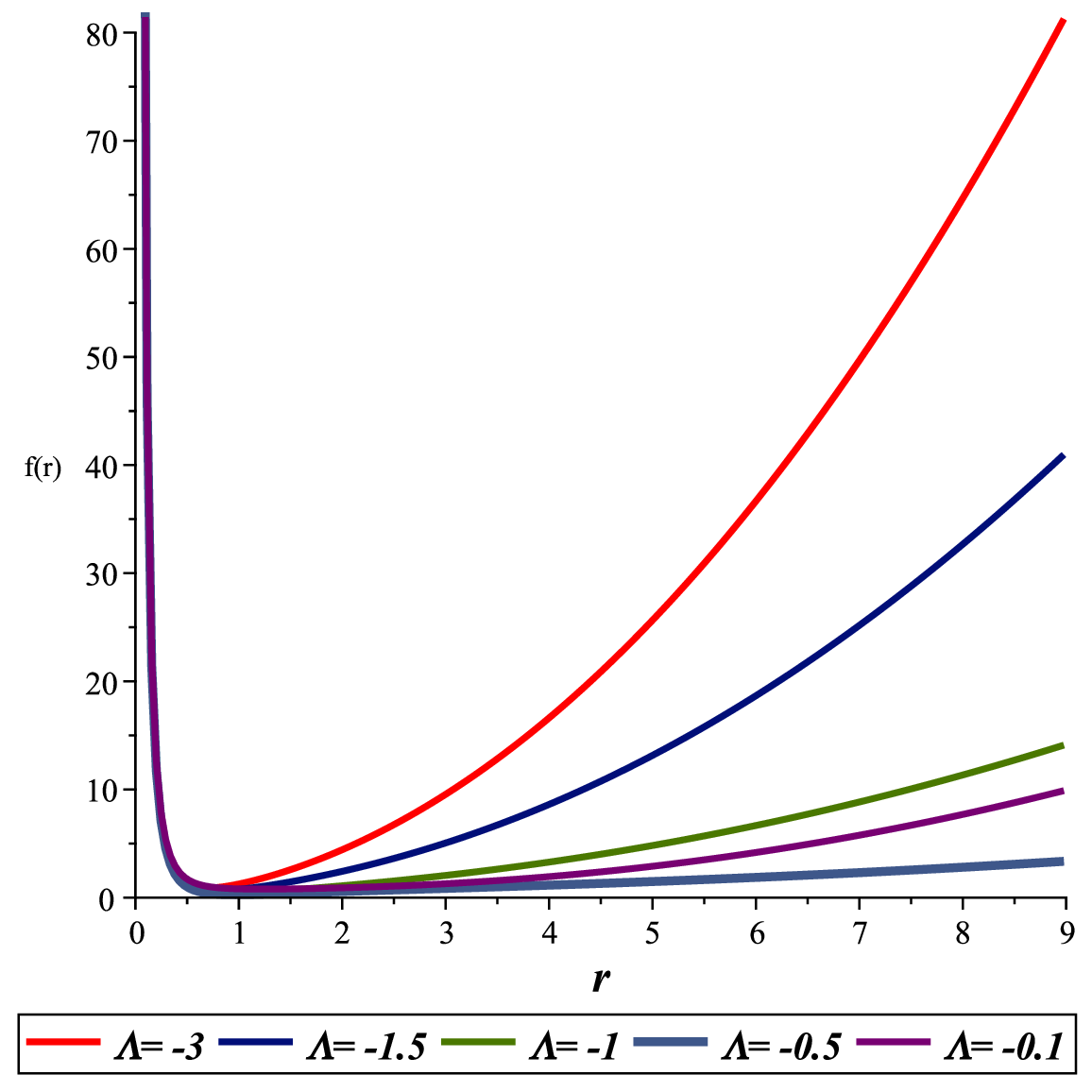}\label{11b}}\\
\subfigure[]{\includegraphics[width=1\linewidth,height=3.7cm]{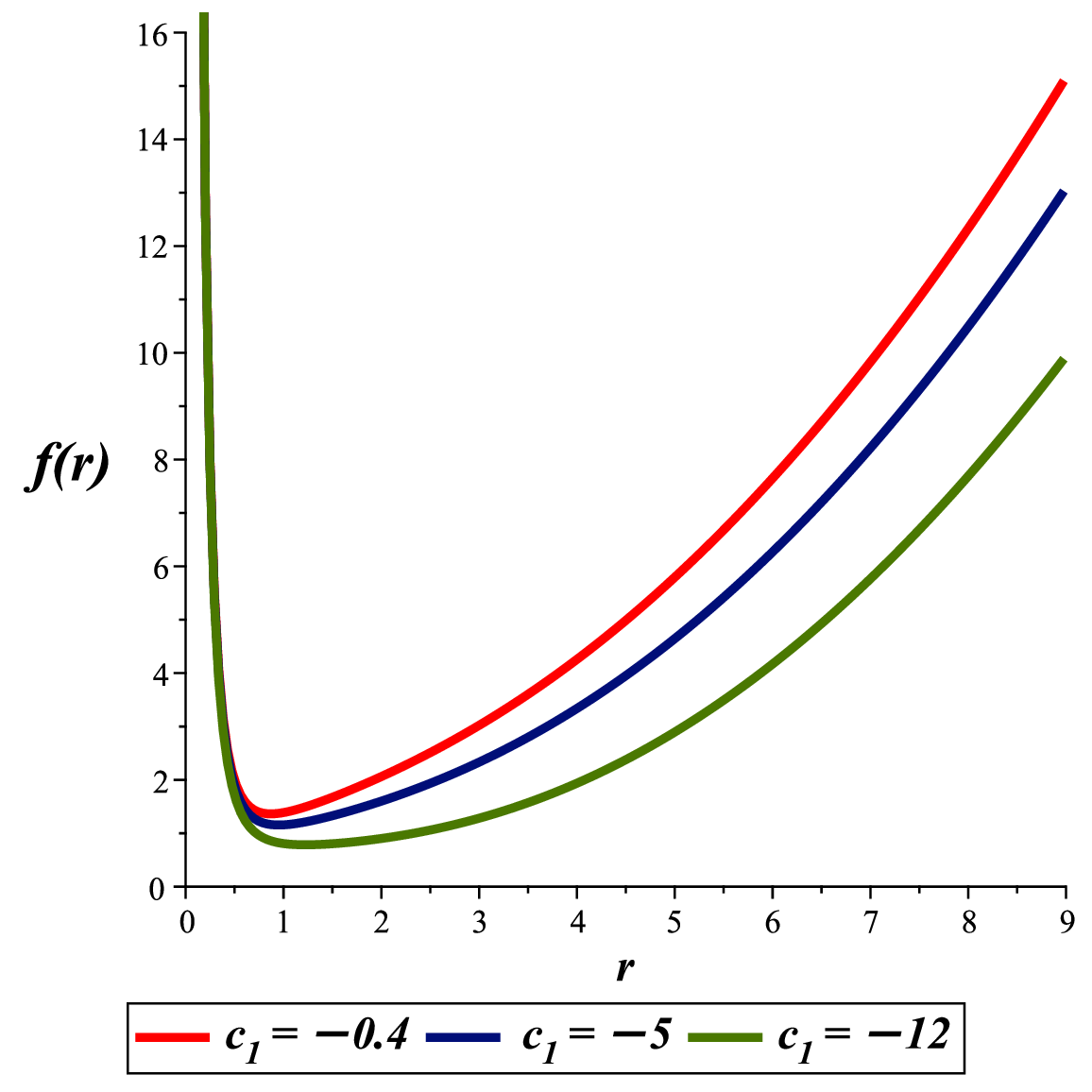}\label{11c}}
\caption{\small{Metric function for the ModMax-dRGT-like black hole: (a) with $M = 1.5, \Lambda = -0.5, \gamma = 0.3, q = 1, C = 0.4, c_{1} = -12, c_{2} = 25$ and varying $m_{g}$; (b) with $M = 1.5, \gamma = 0.3, q = 1, C = 0.4, c_{1} = -12, c_{2} = 25, m_{g} = 0.5$ and varying $\Lambda$; (c) with $M = 1.5, \Lambda = -0.5, \gamma = 0.3, q = 1, C = 0.4, c_{2} = 25, m_{g} = 0.5$ and varying $c_{1}$.}}
\label{App-11}
\end{center}
\end{figure}

\begin{center}
\textbf{1. Extremal and Sub-critical Regime ($\gamma < 0.6686$)}
\end{center}
As depicted in Fig. (\ref{App-11}), the spacetime exhibits horizon structures corresponding to black hole behavior only when the graviton mass approaches the characteristic scale of the model. Otherwise, parametric variations leave the system in a singular state without an event horizon.

\begin{figure}[H]
\begin{center}
\subfigure[]{\includegraphics[width=1\linewidth,height=3.7cm]{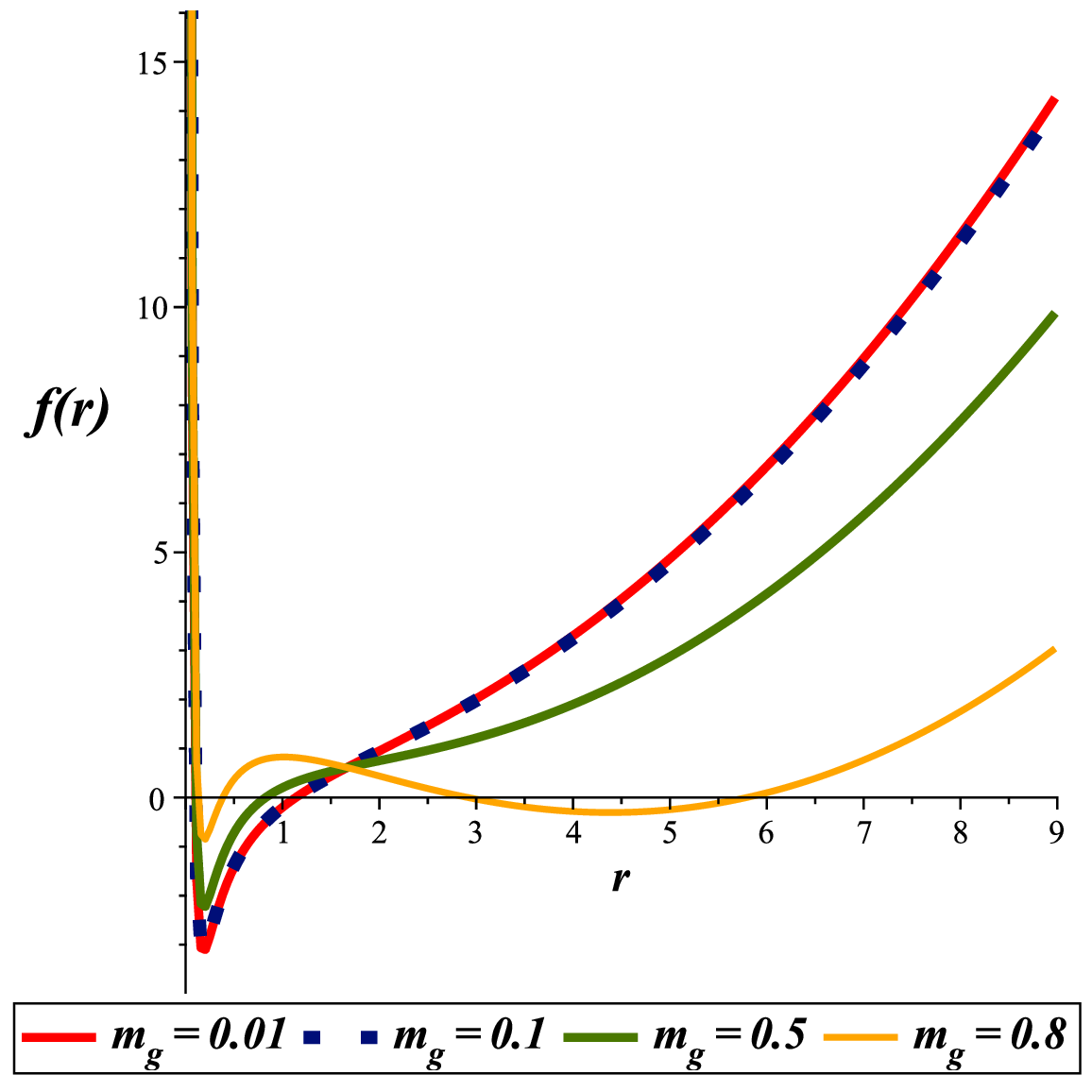}\label{12a}}\\
\subfigure[]{\includegraphics[width=1\linewidth,height=3.7cm]{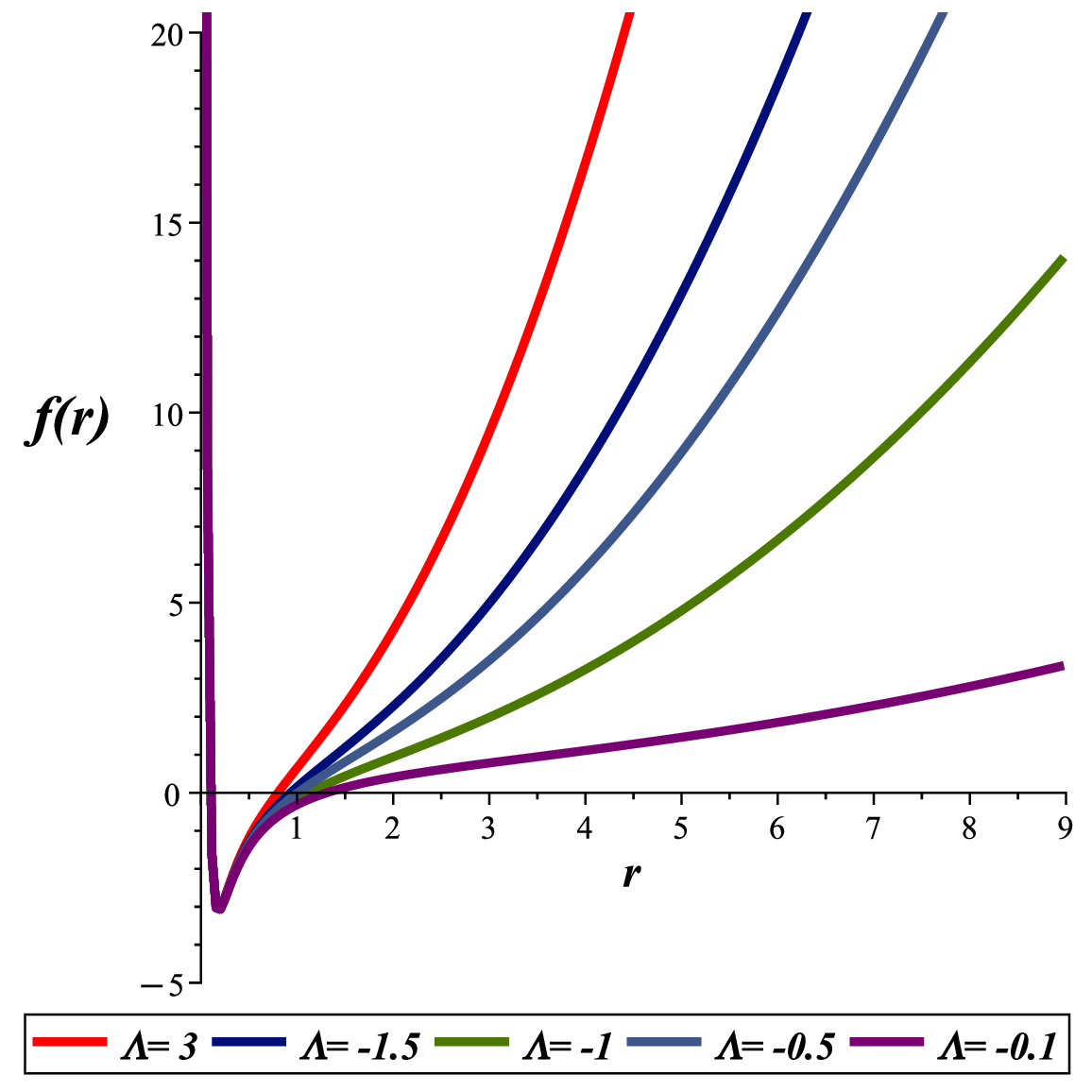}\label{12b}}\\
\subfigure[]{\includegraphics[width=1\linewidth,height=3.7cm]{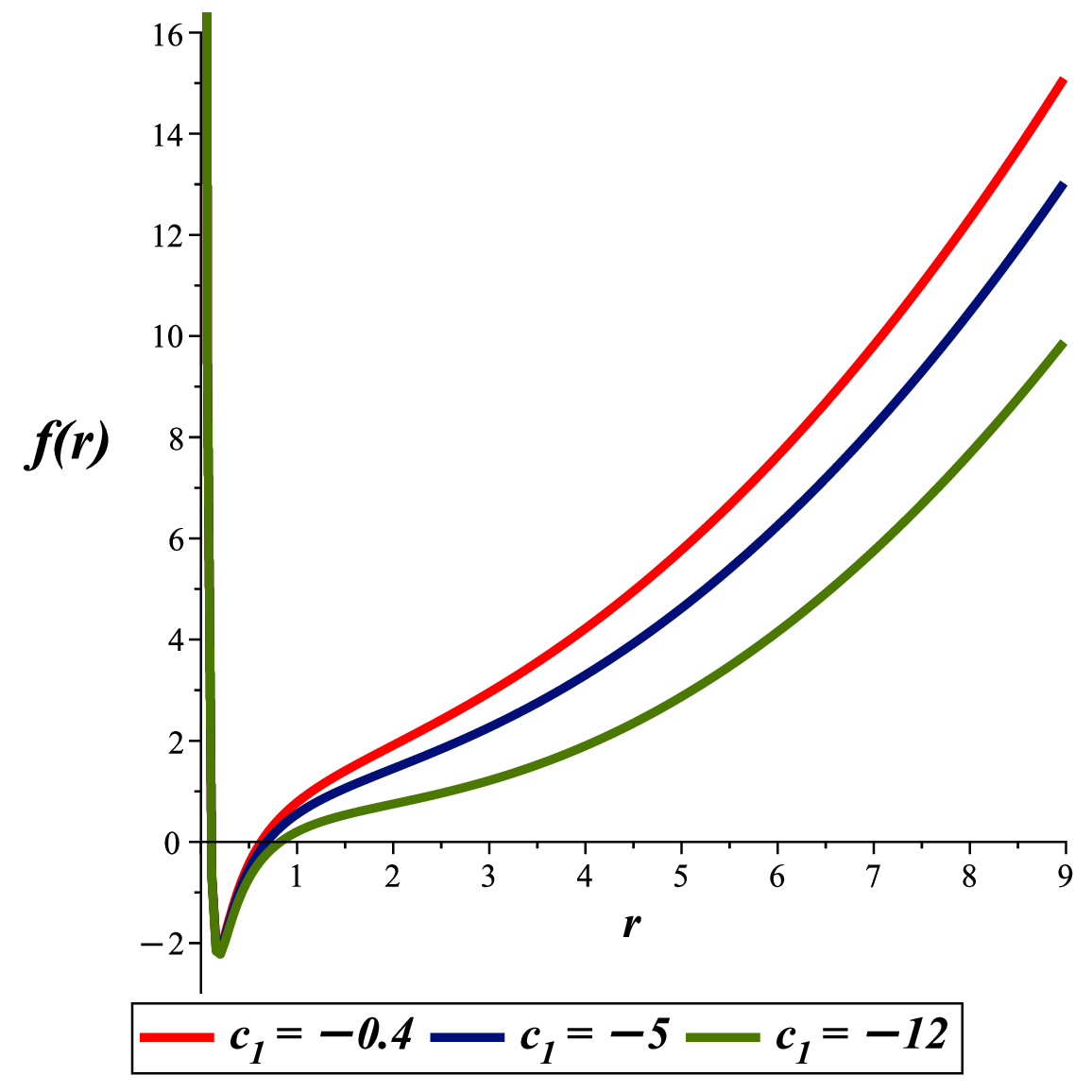}\label{12c}}
\caption{\small{Metric function for the ModMax-dRGT-like black hole: (a) with $M = 1.5, \Lambda = -0.5, \gamma = 2, q = 1, C = 0.4, c_{1} = -12, c_{2} = 25$ and varying $m_{g}$; (b) with $M = 1.5, \gamma = 2, q = 1, C = 0.4, c_{1} = -12, c_{2} = 25, m_{g} = 0.1$ and varying $\Lambda$; (c) with $M = 1.5, \Lambda = -0.5, \gamma = 2, q = 1, C = 0.4, c_{2} = 25, m_{g} = 0.1$ and varying $c_{1}$.}}
\label{App-12}
\end{center}
\end{figure}

\begin{center}
\textbf{2. Black Hole Regime ($\gamma > 0.6686$)}
\end{center}
As illustrated in Fig.~(\ref{App-12}), for the $\gamma=2$ configuration, variations in $m_g$, $\Lambda$, and $c_1$ affect the metric function and its horizon structure, with the remaining parameters fixed in each panel.

\end{document}